\documentclass[12pt]{article}

\usepackage{graphics,epsfig}
\usepackage{amsbsy}
\usepackage{amsfonts}
\usepackage{amsmath}
\usepackage{graphicx}

\def\id{\protect{{1 \kern-.28em {\rm l}}}}

\def\k{\kappa}

\def \cC {{\cal C}}

\def\p{{\partial}}
\def\nn{\nonumber}

\def \bT{{\bar T}}

\def\dalemb#1#2{{\vbox{\hrule height .#2pt
        \hbox{\vrule width.#2pt height#1pt \kern#1pt
                \vrule width.#2pt}
        \hrule height.#2pt}}}

\let\a=\alpha \let\b=\beta \let\g=\gamma \let\d=\delta \let\e=\epsilon
\let\z=\zeta  \let\th=\theta  \let\k=\kappa
\let\l=\lambda \let\m=\mu \let\n=\nu \let\x=\xi \let\p=\pi 
\let\s=\sigma \let\t=\tau   \let\c=\chi 
\let\vp=\varphi \let\vep=\varepsilon
\let\w=\omega      \let\G=\Gamma \let\D=\Delta \let\Th=\Theta \let\L=\Lambda
 \let\P=\Pi \let\S=\Sigma  
\let\C=\Chi \let\W=\Omega
\let\la=\label \let\ci=\cite 
  
\def\nn{\nonumber} \def\bd{\begin{document}} \def\ed{\end{document}}
\def\ds{\documentstyle} \let\fr=\frac \let\bl=\bigl \let\br=\bigr
\let\Br=\Bigr \let\Bl=\Bigl
\let\bm=\bibitem
\let\na=\nabla
\def\tU{{\widetilde U}}
\let\pa=\partial \let\ov=\overline
\def\ie{{\it i.e.\ }}
\newcommand{\be}{\begin{equation}}
\newcommand{\ee}{\end{equation}}
\def\ba{\begin{array}}
\def\ea{\end{array}}
\def\ft#1#2{{\textstyle{{\scriptstyle #1}\over {\scriptstyle #2}}}}
\def\fft#1#2{{#1 \over #2}}
\def\F#1#2{{ F_{#1}^{(#2)} }}
\def\cF#1#2{{ {\cal F}_{#1}^{(#2)} }}

\def\={\, =\, }
\def\+{\, +\, }
\def\-{\, -\, }

\def\R{{\bf R}}
\def\sst#1{{\scriptscriptstyle #1}}
\def\oneone{\rlap 1\mkern4mu{\rm l}}
\def\e7{E_{7(+7)}}
\def\td{\tilde}
\def\wtd{\widetilde}
\def\im{{\rm i}}
\newcommand{\ho}[1]{$\, ^{#1}$}
\newcommand{\hoch}[1]{$\, ^{#1}$}
\newcommand{\bea}{\begin{eqnarray}}
\newcommand{\eea}{\end{eqnarray}}
\newcommand{\ra}{\rightarrow}
\newcommand{\lra}{\longrightarrow}
\newcommand{\Lra}{\Leftrightarrow}
\newcommand{\ap}{\alpha^\prime}
\newcommand{\bp}{\tilde \beta^\prime}
\newcommand{\cB}{{\cal B}}
\newcommand{\cO}{{\cal O}}
\newcommand{\vecx}{\vec{x}}
\newcommand{\vecy}{\vec{y}}
\newcommand{\vecp}{\vec{p}}
\newcommand{\vecq}{\vec{q}}
\newcommand{\tr}{{\rm tr} }
\newcommand{\Tr}{{\rm Tr} }

\newcommand{\cL}{{\cal L}}
\newcommand{\cA}{{\cal A}}
\newcommand{\cD}{{\cal D}}
\def\sst#1{{\scriptscriptstyle #1}}

\def\ve{\varepsilon}
\def\vf{\varphi}
\def\F{\Phi}
\def\wg{\wedge}

\def \foot {\footnote}
\def \bi{\bibitem}

\def \tr {{\rm tr}}
\def \ha {{1 \over 2}}
\def \td {\tilde}
\def \ci{\cite}
\def \N {{\mathcal N}}
\def \ww {\Omega}
\def \const {{\rm const}}
\def \ss {\sum_{i=1}^3 }
\def \t {\tau}
\def\S{{\mathcal S} }
\def \nn {\nu}
\def \XX {{\rm X}}

\def \lra {\leftrightarrow}
\def \vom {{\bar \omega}}
\def \E {{\mathcal  E}} \def \J {{\mathcal  J}}
\def \YY {{\rm Y}}

\def \d {\del}
\def \rJ {{J}}
\def \sms {sigma models\ }
\def \sm {sigma model\ }
\def \L {\Lambda}
\def \gl {\ell}
\def \tr {{\rm tr\ }}
\def\z{\zeta}
\def\zi{\zeta_1}
\def\zii{\zeta_2}
\def\K{\mbox{K}}
\def\eE{\mbox{E}}   \def \vt {\vartheta}
\def \vr {\varrho}
\def \wup {w}

\def\dg{\dagger}
\def\a{\alpha}
\def\b{\beta}
\def\e{\varepsilon}
\def\p{\phi}
\def\ap{\alpha^\prime}
\def\I{{\cal I}}

\def\R{{\bf R}}
\def\Z{{\bf Z}}
\def\C{{\bf C}}
\def\P{{\bf P}}
\def\xb{{\bar X}}
\def\Tr{{\rm  Tr}}
\def\tr{{\rm  tr}}

\def \del{\partial}
\def \a {\alpha}
\def \aa {{\a'}}
\def\g{\gamma}
\def\s{\sigma}
\def\z{\zeta}
\def\zi{\zeta_1}
\def\zii{\zeta_2}
\def\ov{\over}

\def\I{{\cal I}}
\def\J{{\mathcal J}}
\def \ok {{1\ov \k}}
\def\LL{{\mathcal L }}
\def \jL {{J}}
\def \om {\omega}
\def \cL {{\mathcal L}} \def \cH {{\mathcal H}}
\def\E{{\mathcal E}}
\def\w{\omega}
\def\b{\beta}
\def\l{\lambda}
\def\eps{\epsilon}
\def\vep{\varepsilon}
\def \De {{\mathcal D}}

 \def \cV {{\cal V}}
\def  \Jt {  {J}_{\rm tot}    }

\def \k {\kappa}
\def\foot{\footnote}
\def \four{{\textstyle {1\ov 4}}}
 \def \third { \textstyle {1\ov 3
}}
\def\det{\hbox{det}}
\def \ci {\cite}

\def \foot {\footnote}
\def \bi{\bibitem}

\def \tr {{\rm tr}}
\def \ha {{1 \over 2}}
\def \tid {\tilde}
\def \vv {{\rm v}}
\def \tl {{\tilde \l}}
\def \XX {{\rm X}}
\def \ta {{\tilde \a}}
\def \fo { {1\ov 4}}
\def \ep {\epsilon}
\def \inti {{\int^{2\pi}_0 {d \sigma \ov 2 \pi}}}

\def \d {\partial}
\def \K {{\rm S}}
\def \el {\ell}
\def \Tr {{\rm Tr}}
\def \P {\Phi}
\def \l  {\lambda}
\def \tl {{\tilde \l}}
\def \bl {{\tilde \l}}
\def \const {{\rm const}}
\def \V {v}

\def \bv {v^*}
\def \vv {{\rm v}}
\def \LL {{\mathcal L}}
\newcommand{\PV}[1]{P_{\!\!_{V_{#1}}}}

\def \bL {\ell}
\def \M {{\mathcal M}}
\def \N {{\mathcal N}}
\def \S {{\rm S}}
\def \vn {\vec n}
\def \tl {\td \l}
\def \td {\tilde}
\def \Prod {\Pi}
\def \O {{\mathcal O}}
\def \Q {{\rm  Q}}
\def \D {\Delta}
\def \N {{\mathcal N}}
\def\tN{{\tilde N}}

\def \m {\mu}
\def \vs {\vec \s}
\def \ie {i.e.}

\def \cD {{\cal D}}

\def  \le  {\l_{\rm eff}}

\def \rS {{\rm S}}
\def\as{{\a}}
\newcommand{\bra}[1]{\mbox{$\langle #1 |$}}
\newcommand{\ket}[1]{\mbox{$| #1 \rangle$}}

\newcommand{\auth}{AUTHORS}

\def\thb{\bar{\theta}}
\def\Thb{\bar{\Theta}}
\def\barp{\bar{p}}
\def\barq{\bar{q}}
\def\barc{\bar{c}}
\def\bard{\bar{d}}
\def\e{\epsilon}

\def \bi{\bibitem}
\def \la {\label}

\def \l {\lambda}
\def\foot{\footnote}
\def \tl  {{\tilde \l}}
\def \sql {{\sqrt \l}}
\def \adss {$AdS_5 \times S^5$\ }
\newcommand{\rf}[1]{(\ref{#1})}
\def \ov {\over}

\def\th{\theta}
\def\Th{\Theta}
\def\vth{\vartheta}

\def\vth{\vartheta}

\def\ra{\rightarrow}
\def\N{{\cal N}}
\def\F{{\cal F}}
\def\cc{\circ}
\def\eqv{\equiv}

\def\ni{\noindent}
\def \ha{{1\ov 2}}
\def \bw {{\rm w}}

\def\r{{\rm r}}

\def \cT {{\cal T}}
\def \no {\nonumber}

\def \J {\mathcal{J}}
\def \del {\partial}

\def \bps {{\bar \psi}}
\def \sqbl {\sqrt{\bar \lambda}}

\def\dF{\dot{F}}
\def\dG{\dot{G}}
\def\df{\dot{f}}
\def \E {{\cal E}}

\def \S {{\cal S}}
\def \J {{\cal J}}

\def\ms{\mathcal{S}}
\def\mj{\mathcal{J}}
\def\soj{\fr{\ms}{\mj}}
\def \R {{\bf R}}
\def \om {\omega}
\def \tH {\widetilde H}
\def \bE {\bar E}
\def \x {{\cal X}}

\def \hV {{\hat V}}

 \def \bb {\bar \beta}
\def \W {{\cal E}}

\def \bi{\bibitem}
\def \la {\label}

\def \l {\lambda}
\def\foot{\footnote}
\def \tl  {{\tilde \l}}
\def \sql {{\sqrt \l}}
\def \sqtl {{\sqrt {\tilde \l}}}

\def \HH {{\rm E}}
\def \cS {{\cal S}}
\def \cL {{\cal L}}

\def \adss {$AdS_5 \times S^5$\ }

\def \D {\Delta}
\def \thet {\theta}
 \def \t {\tau}
 \def \p {\phi}
 \def \r {\rho}
 \def \rN {{\rm N}}
 \def\tw{{\tilde w}}
 \def\hJ{{J}}
 \def\hw{{w}}
 \def\hl{{\lambda}}
 \def\hth{{\theta}}
 \def\NN{{\cal N}}
 \def \bv {{ \bar w}}
\def \vn {{\vec n}}
\newcommand{\sfrac}[2]{{\textstyle\frac{#1}{#2}}}
\def \bl {{ \bar \lambda}}
\def \bp {{\bar p}}
\def \bu {{\bar u}}
\def \sha {\sfrac{1}{2}}
\def \w {\omega}
\def \ov {\over}
\def \vl { \vec \ell}
\def \varpi {{\rm w}}
\def \OO {{\cal O}}
\def \bG {\bar \G}

\def \c {\gamma}

\def \ss {{\rm s}}

\def \KK {{\rm K}}

\def \ve {\varepsilon}
\def \pa{\partial}
\def \I {{\cal I}}
\def \LL {{\cal L}}
\def \ep {\epsilon}
\def \R {{\rm R}}
\def \tilt {{\tilde t}}

\def\pic #1#2{\hbox{\lower#1pt\hbox{~\mbox{\epsfxsize=20truemm \epsffile{#2}}}}}

\def\pic #1#2#3{\hbox{\lower#1pt\hbox{~\mbox{\includegraphics[scale=#3]{#2}}}}}

\def \bt {\bar\theta}
\def \te {\theta}

\def \cc {{\rm f}}
\def \d {\delta}

\def \cL {{\cal L}}
\def \S  {{\rm S}}
\def \pp {{q}}
\def \vt {\vartheta}
\def \mm {{\cal  \ell}}
\def \Z {{\cal Z}}
\def \pa {\partial}
\def \C {{\cal C}}
\def \be {\bea}
\def \ee {\eea}
\def \c {\gamma}  \def \d {\delta}

\begin{document}
\overfullrule=0pt
\parskip=2pt
\parindent=12pt
\headheight=0in \headsep=0in \topmargin=0in \oddsidemargin=0in


\rightline{Imperial-TP-AT-2007-1}

\

\

\begin{center}

{\Large\bf
Two-loop world-sheet  corrections\\
\vspace{0.2cm}
in $AdS_5 \times S^5$ superstring
   }

 \vspace{.5cm} { R. Roiban$^{a,}$\footnote{radu@phys.psu.edu},
  A. Tirziu$^{b,}$\footnote{tirziu@mps.ohio-state.edu}
 and A.A.
 Tseytlin$^{c,}$\footnote{Also at
 Lebedev  Institute, Moscow.
  tseytlin@imperial.ac.uk
 }}\\
 \vskip 0.3cm

{\em $^{a}$Department of Physics, The Pennsylvania  State University,\\
University Park, PA 16802 , USA\\
$^{b}$Department of Physics, The Ohio State University,\\
Columbus, OH 43210, USA\\
\vskip 0.08cm $^{c}$  Blackett Laboratory, Imperial College,
London SW7 2AZ, U.K. }

\end{center}

 \begin{abstract}

We initiate the  computation of the
2-loop quantum $AdS_5 \times  S^5$ string corrections on the example
of a certain string configuration in $S^5$
related by an analytic continuation to a folded rotating
string  in $AdS_5$ in the ``long string'' limit. The 2-loop term
in  the energy of the  latter should represent
the  subleading strong-coupling correction to the cusp
anomalous dimension
and thus provide a further check of
recent conjectures about the exact structure of the Bethe ansatz  underlying
the AdS/CFT duality. We use the conformal gauge and several choices of
the $\kappa$-symmetry gauge. While we are unable to verify the 
cancellation of 2d UV  divergences 
we compute the bosonic contribution to the effective action
and also determine the non-trivial finite part of the  fermionic contribution.
Both the bosonic and the fermionic contributions to the string 
energy   happen to be  proportional to the Catalan's constant. 
The resulting value for 2-loop superstring prediction for the 
subleading coefficient $a_2$  in the   scaling function  matches 
the numerical value found in hep-th/0611135  from the BES equation.


\end{abstract}
\newpage

\renewcommand{\theequation}{1.\arabic{equation}}
 \setcounter{equation}{0}

\setcounter{equation}{0} \setcounter{footnote}{0}
\setcounter{section}{0}

\section{Introduction}

To demonstrate the  AdS/CFT duality one is to establish
a direct equivalence  between the spectrum of  the $\N=4$ SYM
 dilatation
operator and the spectrum of quantum string energies in \adss.
There are strong indications  that both spectra are indeed
described by solutions of certain
Bethe ans\"atze
(for a recent  review and some references see, e.g., \ci{bes}).

While the  gauge-theory  side of the  duality has
standard definition  at weak-coupling, the presence of the RR background
supporting \adss
requires  that the formulation of the dual string theory  should be based
on the manifestly-supersymmetric  Green-Schwarz  approach \ci{gs,howe}
which leads to a complicated-looking non-linear action \ci{mt,kal,mett}.

The quantization of this action is straightforward at leading
semiclassical (1-loop) order by expanding near a non-trivial classical
string configuration and fixing an appropriate $\k$-symmetry gauge
(see, e.g., \ci{kalt,dgt,mets,ft1}).  This allowed one to compute
1-loop string corrections to energies of various classical solutions
in \adss \ci{ft1,ft2,ft3,fpt,ptt,ftt}, and these explicit results
played a key role in checking the AdS/CFT duality and, in particular,
in recent progress in fixing the structure of the ``string''
(strong-coupling) form \ci{afs} of the Bethe ansatz
\ci{bt,sz,hl,krj} which led to the exact   
expressions in \ci{bhl,bes}.\foot{An additional
input was the assumption of crossing symmetry \ci{jan,afr}.}

To provide further important checks of the conjectured form of the
Bethe ansatz for the gauge/string spectrum it is crucial to learn how
to extend the 1-loop computations of \ci{ft1}--\ci{ftt} beyond the
1-loop level.  Here, however, one faces an apparent problem: the
curved-space GS action expanded near a string background that provides
the fermions with a non-trivial propagator is formally
non-renormalizable beyond one loop.  While the original string action
has no dimensional parameters and both the bosonic and the fermionic
fields in it are dimensionless, when expanding near a non-trivial
background one effectively changes the dimension of fermions to
canonical Dirac field dimension (1/2) in 2 dimensions. The effective
dimensional scale is introduced by the derivative of the bosonic
string background, leading to non-renormalizable couplings (and thus
to higher power divergences).\foot{It is sometimes said that one
cannot quantize GS action since fermions $\theta$ ``do not have a
propagator''. This is somewhat a misleading statement.  The
quantization of the \adss action is formally well-defined as soon as
one chooses a non-trivial bosonic background near which one can expand
the action (and fixes a proper $\k$-symmetry gauge).  There is an
analogy with the
quantization of Einstein's theory: unless one chooses a non-zero
background metric the metric fluctuations do not have a propagator
term -- the Einstein action is non-polynomial in the metric.
Specifying a background metric introduces a dimensional coupling and
also spontaneously breaks the diffeomorphism invariance of the
Einstein action; it can be formally maintained using the background
field method in which the background metric is also transforming
(provided one uses a background-covariant gauge).  Similar approach
can be followed for the GS string. In most practical applications
(see, e.g., \ci{ft1,ft3}) one needs to expand near a specific
background which spontaneously breaks symmetries of the original
action, just as in a generic case of the semiclassical expansion near
a solitonic solution.  }

 This problem did not seem to be appreciated in early studies of
 quantum GS action which were restricted to 1-loop order \ci{gri}, but
 it was recently emphasized in \ci{pola}, where it was suggested that
 it may be possible to resolve it in a special ``light-cone''-type
 gauge.  On general grounds, one should not expect any meaningful
 results to depend on a particular gauge choice, but the formulation
 of quantum theory may look simpler in a gauge where the action has
 less non-linear form (e.g.  being quadratic in l.c. gauge in flat
 space).\foot{A possible alternative is to use the Berkovits version
 of the \adss GS action \ci{ber} that has a non-degenerate fermionic
 quadratic term from the start and formally defines a renormalizable
 theory.  However, the formulation of the theory (using BRST symmetry
 as a basic principle) is somewhat {\it ad hoc} and is not completely
 free of ambiguities (in particular, in the definition of the ghost
 path integral measure). To see if this formulation is of practical
 use for addressing the issues discussed here
%
%
it would
be important to first reproduce  the results of the
1-loop GS computations in  \ci{ft1}--\ci{ftt}
by starting with the  Berkovits action.}

On general grounds, one should expect the GS action to make sense at
the quantum level only if it happens to be UV finite: this is required
by its basic gauge symmetry -- the $\k$-symmetry.  The key technical
issue is how to formulate the quantum theory (i.e. make a choice of a
regularization, measure, etc.)  in a way that is indeed consistent
with the preservation of the classical symmetries at the quantum
level.
%
%
\foot{By this we mean, in particular,  that the $\kappa$-symmetry does
not develop
anomalies, i.e. anomalies cancel.  The usual quantization schemes
specify a regulator that preserves as many symmetries as
possible. Anomalies may arise, however, if a symmetry is broken by the
regulator. A formal argument for finiteness of the \adss action
\ci{mt} constructed by analogy with the one for the WZW theory runs as
follows: (i) the ``kinetic'' term in the action is protected by global
symmetry (as for, e.g., $SO(n)$ coset sigma model) and can thus be
renormalised only by an overall factor; (ii) the coefficient of the WZ
term in the action of \ci{mt} cannot be renormalised (for a symmetric
supercoset the analog of the field strength of the $B_{mn}$ coupling
is covariantly constant; alternatively, the WZ term has a 3d
representation that is not possible for local covariant counterterms);
(iii) the $\k$-symmetry relates the coefficients of the WZ and the
``kinetic'' terms, thus precluding any renormalization of the latter.
This argument is very formal since it assumes that both global
supercoset symmetry and the $\k$-symmetry are actually preserved at
the quantum level.  The main issue is how to formulate the quantum
theory explicitly so that these conditions are indeed met.}


\bigskip

Our aim here will be  to begin   the investigation of the
quantum \adss  string theory beyond the  1-loop order by attempting to compute a
2-loop correction to the string world-sheet effective action
in  a  particular   string background.
This background appears to be one of the   simplest possible non-trivial
choices, making the 2-loop computation
%
%
tractable.
It  may be viewed as   a particular limit of the  circular string
solution with two equal $SO(6)$ spins  \ci{ft2,art}
and  is an example of a  ``homogeneous'' spinning string solution  for which
the only non-vanishing string coordinates are  isometric angles of \adss
 which are linear in    string world-sheet coordinates    $\tau$ and $\sigma$.
 This choice
  is  special in that, when expanded near it,  the \adss string
 Lagrangian  has {\it  constant}  ($\tau,\sigma$ independent)
 coefficients and thus the computation  of quantum corrections simplifies
considerably.
%
%
 An apparent  problem, however, is that the simplest
 spinning string solution with two equal $SO(6)$ spins  \ci{ft2,art}
 is unstable, and that seems to lead to potential  problems in
 trying to compute the  2-loop  correction to its 
energy.\foot{A similar ``homogeneous''
 circular string solution with one spin
  in $AdS_5$ and one  in $S^5$
 {\it is} stable, but the corresponding fluctuation 
spectrum (and thus the  propagator)
 is much more involved \ci{ptt},
 substantially  complicating the problem of computing the 
2-loop correction.} One  may
%
%
avoid  this  instability  by  a
  formal analytic continuation
  in the winding number $m$, i.e. by  taking it  less
   than one or even purely  imaginary.

Remarkably, there is also  another important reason to
study quantum corrections  to the  energy of the circular 2-spin $S^5$ solution
with an   imaginary  winding parameter.
As was noticed   recently  \ci{ftt} (using an  earlier  observation in \ci{bfst}),
  this solution
is related  by a formal
 analytic continuation to a  ``long-string'' limit of  the
 folded  string  rotating in $AdS_5$
with spin $S$  and  also orbiting along big circle of $S^5$ with spin $J$.
The energy of the $S \gg J $  string  \ci{gkp,ft1}
goes  as $ S + a_0 \sql  \ln {S \ov J}$ in the  ``long-string'' limit, and it
played a key role  in recent
 discussions of the AdS/CFT correspondence in the
$SL(2)$ sector \ci{bes,bern,kleb,lip,klebb,serb}.

Let us start with introducing the relevant string background
and reviewing the form of the 1-loop correction to its  energy.

\subsection{String background and
strong-coupling expansion of  minimal  twist anomalous
dimension}

According to  \ci{gkp,ft1}
the classical energy  of a folded rotating string
in $AdS_5\times S^5 $ which should be dual to a minimal twist operator in planar
$\N=4 $ SYM theory
 scales for $ S \gg J$  as\foot{More precisely, one is to assume that
  $\ln { S\ov J}  \gg {J\ov \sql } $; we also omit
a term linear in $J$ on the r.h.s.}
 \be \la{sca}
  E=S+ f(\l)  \ln S +   ... \ .   \ee
For small $\l$  the    function $f(\l)$  should
   have  the standard
 perturbative gauge theory   expansion
$f(\l)= k_1 \l + k_2 \l^2 + ...$   while for large $\l$
it should have  perturbative string
theory expansion
\be \la{uuu}
f(\l)= a_0  {\sql}    + a_1   + { a_2\ov  \sql}   + ...    \ , \ \ \ \  \ \ \
a_0 = { 1 \ov \pi} \ , \ \ \   a_1 = - { 3 \ln 2  \ov \pi }  \ . \ee
The leading  strong-coupling   coefficients $a_0$ \ci{gkp}  and $a_1$  \ci{ft1}
 were found
to be  in perfect agreement  \ci{bes,bern,kleb}  with the  prediction of
 the integral equation for the
minimal twist anomalous dimension
 as  extracted from the
weak-coupling Bethe ansatz suggested in \ci{bes}.\foot{
This may not be totally surprising  since the 1-loop dressing phase in the strong
coupling Bethe ansatz   was extracted \ci{hl} from other 1-loop string
results; nevertheless, it
provides a  non-trivial
check   of the analytic continuation prescription suggested in \ci{bes},
as it implies the existence of a single function with correct  weak-coupling
and strong-coupling
limits.}

It is  obviously  important  to  compute the value  of the  subleading coefficient
$a_2$  directly as the  {\it two-loop}   correction in the \adss
string theory. It can then be compared with a
prediction of \ci{kleb}  obtained numerically
 from the  strong-coupling expansion
of the  solution  of the  integral equation
of \ci{bes}:\foot{The strong-coupling expansion of this integrals
equation appears to be subtle
\ci{klebb}.
 It would   be important to obtain
the expressions for the strong-coupling coefficients
 $a_1, a_2, ...$ analytically.
Note  that the coupling $g$ used in \ci{bes,kleb}
 is related to $\l$ used here by
$g= { \sql \ov 4 \pi }$.}
\be  \la{kkkk}
 a_2 \approx   - 0.29154\pm 0.0013  \ .
\ee
In general, computing quantum corrections to the energy of the folded string
solution in $AdS$   \ci{veg,gkp} is very complicated
due to the  non-trivial $\s$-dependent form of this  configuration.
 However, as was realized  in \ci{ft1,ftt}
to extract the leading large spin ${S\ov \sql }\gg 1 $ behaviour of the
 energy it is sufficient to consider
the  ``long string''   approximation
 in which the folded string solution simplifies,  becoming  effectively
 ``homogeneous''.
  Viewed as a string configuration in  $AdS_3\times S^1 $
 (where $S^1$ is from $S^5$)
 with the metric
 \be\la{ds}
  ds^2 = d\r^2 - \cosh^2 \r\   dt^2 + \sinh^2 \r\   d \theta^2 + d \p^2 \ee
 it is then  approximated  (in conformal gauge) by
 \be
 \la{ggg}
 t= \k \tau,\ \ \ \ \
  \ \theta \approx \k \tau,\ \ \ \ \
   \ \r\approx \mm \sigma,\ \ \ \ \
    \ \p =\nu \tau,\ \ \ \  \
     \mm \equiv   \sqrt{ \k^2 - \nu^2} \ ,  \ee
     where $S$ is related to $\k$ and $J = \sql \nu$.
 The relevant   limit  we are interested in is
 \be \la{sc}
 \k
  \gg 1   \ , \ \ \ \ \ \ \ \ \ \ { \nu  \ov  \k }= {\rm fixed}
 \ll  1 \
  \ee
  which is sufficient   for computing
  the coefficient of the leading $\ln S $ term in the energy.


The above configuration \rf{ggg}
is related \ci{ftt}
 by a formal analytic continuation \ci{bfst} to
the $J_1=J_2$ circular string solution  in $R_t \times S^3$ part of \adss
\be \la{sss}
ds^2 =- dt'^2 +  d\psi^2 + \cos^2 \psi\ d\p^2_2 + \sin^2 \psi\ d \p^2_3
\ee
 taken  in its form given in \ci{ft2} ($J_1=J_2 = \sql w$):
  \be \la{rr}
  t'= \k' \tau,\ \ \ \  \ \ \ \ \psi = m \s,\  \ \ \ \ \
  \ \ \  \p_2= \p_3 = w \tau, \
\ \ \ \ \
    w= \sqrt{ \k'^2 - m^2}  \ . \ee
  Under the continuation $t \to \  \p_2, \
    \rho \to i \psi, \ \p \to \p_3, \ \p \to t'$ 
    one effectively interchanges $AdS_5$ with $S^5$, and  to make it an equivalence transformation
    one is  also  to
    change the overall sign of the  string
    action which can be implemented as a formal inversion of the sign of the 
    coefficient in front of the action, i.e.
    \be \sql \ \to \ - \sql  \la{cha} \ . \ee
    The 
     parameters  of the two solutions are related as follows:
\be \la{rrr}
  \kappa'= \nu\  , \ \ \ \ \ \  \  m= i \mm = i\sqrt{ \k^2 - \nu^2}\ , \ \ \ \ \ \ \ \
 w= \k \ .
\ee
The  values of the classical string  action evaluated on \rf{ggg} and on \rf{rr} (or \rf{rre} below) 
 then agree   provided  we also 
do the replacement \rf{cha}.

The quadratic fluctuation action near the above solution \rf{ggg}
will have constant
coefficients after a coordinate  rotation \ci{ft2,ft3}. We may also start directly
with the same background \rf{rr}  in the  equivalent ``rotated''  form given in \ci{art}:
\be
\la{rre}
 \psi = { \pi \ov 4}\ ,\  \ \ \ \ \  \ \ \ \ \
  \ \ \  \p_2= w \tau +m \s\ ,   \ \ \ \ \  \ \ \ \ \
  \p_3= w \tau -  m \s\ .   \ee
Then  all coefficients in the fluctuation Lagrangian will be manifestly
constant.
It is the configuration
  \rf{rre} that will  be our starting point for the  quantum string loop 
  computations here.

Our aim below will  be to compute the 2-loop string correction to the energy of the
circular solution  \rf{rre} assuming the analytic continuation  in $m$ \rf{rrr}
and  the  scaling limit \rf{sc}.
 For simplicity we shall also set $\nu=0$,
 i.e. set the $S^5$ spin of the original  folded string  solution \rf{ggg} 
 to   be  zero
or set $\k' =0$ for the rotating solution in \rf{rrr}:
\be\la{sci}
\k'=\nu=0 \ , \ \ \ \ \ \ \ \ \ \
m = -i \k, \ \ \ \  \ \ \ \ w= \k \ , \ \ \ \ \ \ \ \ \ \ \k \approx { 1 \ov \pi} \ln { S \ov \sql}
\to \infty \ . \ee
In the scaling limit $\k \to \infty$ the world sheet coordinates $\t$ and $\s$  in \rf{rre} can be rescaled
by $\k$  and thus 
 we can 
replace the $R \times S^1$  string world sheet  by
the $R \times R$ one, i.e. 
the  summation over the spatial momentum modes can be  replaced by an integral
 \ci{ft1,ftt}.
 As a result,   the dependence on $\k$ in the effective action
 will factorize.\foot{The argument  about
factorization of $\k$  dependence is strictly
true only  if all divergences  cancel out.
If, e.g., IR divergences were survive  one could get non-analytic $\k^2 \ln \k$
contributions. We  will find that they indeed cancel in the   final result.
The  analytic continuation in the winding $m$
eliminates the tachyonic instability of the circular solution
making the  2d  momentum integrals
better defined in the IR.}

Instead of directly computing the quantum
correction to the energy  of our soliton solution  using  operator methods
we shall find 
the value of the  logarithm of the  string  partition function 
(equal in the present case of  a homogeneous background 
 to the quantum  effective 1-PI action)
 evaluated on  the classical solution.\foot{
Note that  quantum corrections should  not change  the form of the
 classical solution
due to its homogeneous nature. This case is similar  to  the case of  a constant abelian
gauge strength background in  gauge theory.}
We will evaluate  the partition function or the 2d energy by expanding 
near  \rf{rre} in the formal limit $\k'\to 0 $
as a  function of  (in general, complex) argument  $m$ and at the very end set
$m= -i \k$  where  $\k \to \infty$.
The final result should give
us, as it happened  at the tree and the 1-loop level,
the information about the 2-loop correction to the energy of the folded string
solution.\foot{We expect that  the string energy has a 
meaning when considered
as a function of the complex values of its parameters, i.e. that different analytic
continuations in  parameters give  values of the  energy for different physical
configurations. In short, having two classical solutions
related by an analytic continuation in   coordinates and
parameters we shall assume that this relation holds also
at the quantum level. We  cannot of course  consider the rotating solution as
physical in the limit \rf{sci} (e.g., its energy is not defined if $\k'=0$)
but we shall assume that  this limit  of its  energy
defined for complex $\k'$ and $m$    has a meaning of the energy of the folded
solution.}
More precisely, taking into account \rf{cha}, 
 the quantum  \adss superstring  partition function  computed  by expanding near the folded 
 rotated string 
 solution in $AdS_5$ \rf{ggg}   and near the 
 related by the analytic continuation   complex  rotating   string   background 
  in $S^5$ \rf{rre},\rf{sci} 
 should satisfy 
 \be 
 \la{idi}
 \ln Z_{\rm fold. \  AdS_5} (\k; \sql) =  \ln Z_{\rm rot.\  S_5} (w=im=\k; -\sql) 
 \ . \ee
Thus having  found $\ln Z_{\rm rot.\  S_5} (w=im=\k; \sql)= 
\sql c_0 + c_1 + { c_2 \over  \sql } + ... $ we will need 
to reverse the sign of $\sql$ to find the corresponding values of the coefficients 
in the scaling function \rf{uuu}. This will not  change the 1-loop correction but with 
alter the sign of the 2-loop term.


\subsection{One-loop approximation}


As a preparation for the 2-loop computation we are interested in,
 let us    explain  how one can get the same  1-loop correction
as in \ci{ft2,ftt}  by starting with
the 1-loop  effective   action
 $\G_1=-\ln Z_1$ instead of the usual expression for 1-loop energy correction in terms of
 the sum over the characteristic   
  frequencies $\sum_n \omega_n $.\foot{The  two expressions  are of course related in general
by integrating out over $p_0$ component
of the 2d momentum with  the  $i \epsilon$ prescription, but here
in the absence of the UV divergences even a  formal
Euclidean continuation and direct integration over $p_0$  is enough
to obtain  the required  result.}
The expression for the leading term
in the 1-loop correction
to the energy of the folded string found in the scaling limit \rf{sc}  with $\nu= 0$
($\k \to { 1 \ov \pi} \ln S$) is   \ci{ft2,ftt}:\foot{Since $t =\k \tau$,
the space-time
energy is related   \ci{ft1}  to the 2d   energy  by $E= {1 \ov \k} E_{2d}$;
in the limit $\k \to \infty$  the 2d energy  $E_{2d}$
scales as $\k^2$.}
\bea
&& E=  { 1 \ov \k} E_{2d} \ ,\ \ \ \  \ \
E^{(1)}_{2d} =  \pi  \kappa^2 a_1 \ , \ \ \ \ \ \ \
a_1 =  { 1 \ov \pi}  \int^\infty_0{ dp}  \   \omega(p)   \ , \la{ji} \\
&&\omega(p) = \sqrt{ p^2 + 4} + 5 \sqrt{ p^2}  + 2 \sqrt{p^2 + 2} - 8 \sqrt{ p^2 +1}\ . \la{pl}
\eea
Here $\omega(p)$ contains the contributions of 8 bosonic and 8 fermionic fluctuation modes.
The integral over $p$  gives\foot{For the reasons
 mentioned above,
this integral   happens  to be essentially the same
  as
in the case of the 1-loop correction to the energy of the
circular  $J_1=J_2$ string solution in $SU(2)$ sector \ci{ft2,art}
considered in \ci{bt} and  in Appendix C of \ci{mtt2}.}
\be \la{cee} a_1= - {3 \ln  2  \ov \pi}  \ . \ee
We get the same result if we consider instead the expression for the Euclidean
partition function
 and define $E_{2d}$ as the effective action $\G $ divided  over the
2-d  time interval, i.e. at one loop
\be \la{pi}
\G_1 =  - \ln  Z_1  =
 V_2 \int{ d^2 \pp \ov (2 \pi)^2}\   {\Z_1} (\pp^2)
 \ ,
   \ee
 \be \la{lji}
 {\Z_1} (\pp^2)   = \ha  \bigg[  \ln (\pp^2+4) + 5 \ln \pp^2 + 2 \ln (\pp^2+2)
 - 8  \ln (\pp^2+1)\bigg] \ . \ee
Here $V_2  =  L T $ is the 2-d volume which factorises since our
background is homogeneous: the fluctuation Lagrangian has constant coefficients
and is thus
translationally invariant.  We assumed that the original coordinates
$\tau$ and $\s$ were rescaled by $\k$
(this decompactifies the spatial direction in the limit $\k \to \infty$),
so that
\be \la{ken}
 L= 2 \pi \k\ , \ \ \ \ \ \ \ \
 \ T= \k \bT\ , \ \ \ \ \ \ \ \  \ V_2  = L T =
  2 \pi \k^2 \bT \ , \ee
and thus
\be \la{en}
E^{(1)}_{2d} = \bT^{-1} \G_1    \ , \ \ \ \ \ \ \ \ \ \
E_1 = T^{-1} \G_1=   2 \pi \k \int{ d^2 \pp \ov (2 \pi)^2}\   {\Z_1} (\pp^2)
\ . \ee
 The integral over the 2d momentum is defined
 using the Euclidean continuation, i.e. $  \pp^2= q_0^2 + q_1^2$.
 Introducing the
 polar momentum space  coordinates $d^2 \pp = \pp d\pp d \phi$
and integrating  over $\phi$ we end up with
\be \la{uji}
E^{(1)}_{2d} = \ha  \k^2 \int^\infty_0   dv  \    \Z_1 (v)  \ , \ \ \ \ \ \ \ \ \ \ \ \ \ \
v\equiv  \pp^2 \ .
 \ee
This  leads to the same expression for $a_1$
in $ E^{(1)}_{2d}=\pi  \k^2 a_1 $ or in 
\be \la{gao}
\G_1 =  \ha a_1   V_2   \ee
 as in \rf{ji},\rf{cee}.
Note that the classical string  action  evaluated  on  conformal-gauge solution
for the folded string  \rf{ggg} with $\nu=0$
gives $\G_0= { \sql \ov 2 \pi}  V_2 = \ha  \sql a_0 V_2$, 
while in the case of \rf{rre},\rf{sci} we get the opposite sign
$\G_0= - { \sql \ov 2 \pi}  V_2 $;
  the 1-loop correction is the same in both cases, 
in agreement with  \rf{idi}.


\subsection{Structure of the paper}

Below we  shall 
perform  the computation of the 2-loop correction  to the  1-PI
 2d  effective action in the background \rf{rre} in the limit \rf{sci}.
 The two-loop  effective action  or the 
partition function for the folded string solution 
  will again be proportional to the volume  factor, i.e.
\be \la{twol}
 \G= - \ln Z= \ha f(\l) V_2  \ ,\ \ \ \ \ \ \ \ \   
 \G_2 =  \ha { a_2 \ov  \sql}   V_2  \ .  
\ee 
Having found   $\G_2$   for its $S^5$ counterpart   \rf{rre},\rf{sci}  to extract the value of $a_2$ 
in the scaling function \rf{uuu}  
we will  need,     according to \rf{idi},    to  change its overall sign.

 This is a technically involved computation.
 One  issue is the large number of fields (10 bosonic and 32 fermionic)
 implying a large number of 2-loop Feynman graphs  with  non-diagonal propagators.
 Another is the presence of gauge symmetries -- 2d diffeomorphisms (which we will fix by
 the conformal gauge)
  and  the fermionic $\k$-symmetry. The preservation of the latter
 is expected to be  quite subtle  at higher loop orders. The  complicated structure
 of the  GS action  makes the verification  of cancellation of
  UV  divergences (power-like, $\ln^2\L $ and  $\ln\L $ ones)
  non-trivial at the 2-loop order.\foot{ A crucial    issue  is that of  an
invariant UV  regularization.
Since the  \adss action contains the WZ-type term with $\ep^{\a\b}$ tensor
 there are many
analogies  with 2-loop computations in bosonic sigma models
with $B_{mn}$ coupling (see, e.g., \ci{us,et,van}).
  Other technical issues discussed below
  are   cancellation of
  IR divergences (which would be automatically
   absent in the static gauge but
  formally may  remain in the
conformal gauge since some of the modes are massless) and the
 lack of manifest 2d Lorentz invariance (``spontaneously'' broken  beyond quadratic  order
 by our choice of
the background).}

We shall start in section 2  with determining the contribution
 of the  2-loop graphs containing the   bosonic  fluctuations.
   Section 2.1 will review  some general facts  about 2-loop
    renormalization of generic   bosonic 2d sigma model in  dimensional regularization,
     pointing out in
    particular that for symmetric spaces like \adss the corresponding effective action
    does not contain $\ln^2\L \sim { 1\ov \ep^2} $  UV divergences.
    In section 2.2 we shall present the form of the bosonic part of the \adss  action
    expanded  to quartic order near the background \rf{rre},\rf{sci}
    and in section 2.3  will  collect the  expressions for the corresponding 2-loop
    momentum integrals. The explicit results   for the integrals will be presented in
    section 2.4.
    In addition to the standard 2-loop
    logarithmic divergence (that should be cancelled by the fermions)
     we shall find that the  non-trivial finite part of the
    bosonic contribution to the
    2-loop coefficient $a_2$ in \rf{uuu} is  
    proportional to 
    the Catalan's constant  $\KK$. 

In section 3  we shall   summarize the results of the computation of the
2-loop graphs  involving the fermionic   variables  of the \adss   action of \ci{mt}
(the action is reviewed in Appendix A).   We first consider the ``covariant'' $\k$-symmetry gauge
$\theta^1 = k \theta^2$  where   $k$ is a real number. The relevant quartic
 part of the
 superstring action is  given  explicitly
 in Appendix B. As we  explain in  Appendix C,  using
 a similar $k=1$ gauge in the flat-space   GS action  one finds that
 the corresponding  2-loop graphs  vanish in dimensional regularization,
 i.e. the 2-loop term in the flat-space partition function vanishes, in agreement with
 its triviality    in the light-cone gauge.

  Computing the  2-loop graphs resulting from vertices in the  \adss action
 (using a {\tt Mathematica}-based  program   to evaluate several hundred Feynman diagrams)
 we found that their contribution to the  effective action
 contains $\ln^2\L $  UV divergences. Since these   were absent in the bosonic
 contribution,   this  contradicts the expected finiteness of the \adss
 string. Moreover, the coefficients of both the divergent   and the finite
 2-loop  part
 happen to depend on the gauge-fixing parameter $k$.
 This should not happen in an  expansion near a classical solution and
  suggests a potential
 problem in our method of computation   which we are unable to
  resolve at the moment.

 For that reason we also  redo  the computation  in a different
 $\k$-symmetry gauge $\G_+ \theta^I=0$ which is a direct analog of the usual light-cone
 gauge in flat space. The \adss  action in that  gauge is presented in Appendix D.
 There we show also that  expanding the \adss action near  a null geodesic
 that wraps  big circle of $S^5$  and computing the resulting 2-loop  correction
 using  $\G_+ \theta^I=0$  gauge  one finds  that it vanishes, in agreement with
 the BPS   nature of the BMN vacuum state.
 Expanding  near our background  \rf{rre},\rf{sci} using the  light-cone
 $\k$-symmetry gauge we find  again  that the 2-loop
 $\ln^2\Lambda $ divergences  do not cancel.\foot{It
 is hard to attribute this lack of cancellation
to a problem with the quartic fermion terms in  the classical
\adss action as given in Appendix A.
Indeed, these terms provide the four-fermion entries of the tree-level
scattering matrix which have been tested in \cite{KMRZ,AFZ}. Moreover,
these terms
 contribute nontrivially in the near BMN expansion,  leading,
as discussed in Appendix D.1, to the expected cancellation of the 2-loop
correction to  string world-sheet
partition function in the expansion near a null geodesic.}

Despite a problem with non-cancellation of divergences
(due to our lack  of  understanding of how to implement the UV regularization in a way 
consistent with symmetries of the action) a
strong   indication of  consistency of our computation of the {\it finite part}
 is that the  non-trivial finite term in  the 2-loop effective action
 is found to be {\it the same} 
  in the $\G_+ \theta^I=0$  and in the $\theta^1 = \theta^2$
 (i.e. $k=1$) gauges   and 
   like the bosonic contribution,  it is 
   again  proportional to the Catalan's constant $\KK$.
   Moreover, combining  the bosonic   and the fermionic 
   contributions to the 2-loop coefficient
   $a_2$ in \rf{uuu} we  find that 
   \be \la{katl}
   a_2= -{  1 \ov \pi} \KK \approx -0.29156  \ , \ee
   which matches  the numerical result 
   \rf{kkkk} of \ci{kleb}  found from the BES \ci{bes} equation. 
   Remarkably, it  agrees precisely with  
   the exact Catalan constant value of $a_2$ found recently as part of  an impressive 
    complete solution of the BES equation  
   in \ci{bkk}.\foot{We are  very grateful to  G. Korchemsky for sending 
   us the draft of this paper 
   prior to its publication which stimulated us in  debugging the computation of $a_2$ 
   in the original version of our paper.}

    Thus, while we were 
   currently  unable to verify the 2-loop finiteness of the \adss string action,
   an unambiguous conclusion  of  our work  is
  the determination of the  transcendental structure  of the string prediction for 
  $a_2$  and  its agreement with the  result   following from the 
  Bethe ansatz  equation of \ci{bes}.


We make some concluding remarks in section 4.
    Some  details of computation of 2-loop momentum integrals are discussed
    in Appendix E.



\renewcommand{\theequation}{2.\arabic{equation}}
 \setcounter{equation}{0}
\section{Bosonic contribution to the 2-loop effective action}

The  bosonic part of the \adss superstring action
 in the conformal gauge     is simply the direct  sum of the standard 2d
 sigma models on $AdS_5$ and $S^5$.
 The corresponding quantum theories are decoupled  before fermions
  are switched on.  Here we shall consider  the 2-loop contributions
  of the bosonic fluctuations near the string background \rf{rre},\rf{sci}.

\subsection{General remarks  on  bosonic sigma model}

The
 2d sigma model  action is (here we assume a Euclidean world-sheet signature)
\be
I=  { 1  \ov 4 \pi \a' } \int d^2 \s \ G_{\m\n} (x)\ \del^\a x^\m\  \del_\a x^\n
  \ ,  \ee
 where in the case of our interest  $G_{\m\n}$ is the  metric of \adss with  radius
 $a, \   \sql = { a^2 \ov \a'}$.
 If we use an explicit UV cutoff $\L \to \infty$, the non-trivial
  power divergences
 in the  partition function or in the effective action
 (computed by  expanding near a solution of the classical equations of motion)
 should be   cancelled  by  the  covariant measure  contribution
 in
 $Z = \int \prod_\s dx(\s) \sqrt{G(x(\s))} \ e^{-I[x]} $,
 i.e.  by the contribution of the counterterm
  \be \la{mea}
  \Delta I = - \ha  \int  d^2 \s \ \Tr  \ln G(x) \ \delta^{(2)} (\s,\s) \ , \ \ \ \ \ \
  \delta^{(2)} (\s,\s) = { 1\ov 4 \pi} \L^2    \
  \ee
   added to the bare action.\foot{This  covariant measure factor may be understood as appearing from
 a 1-st order ``phase-space'' formulation upon integration over the momenta.
 More generally (in the bosonic sigma model context),
 the quadratic  divergences may be absorbed into renormalization of the
 dimension 0  ``tachyon'' coupling, so in the bosonic string context
  the choice of the measure is like a choice of
 a  bare  value of the tachyon field (see, e.g., \ci{mea}).}

If we use covariance-preserving
 dimensional regularization all power divergences will be absent  automatically,
i.e. the only potential divergences at the 2-loop level will be ${1\ov \ep} \sim \ln \L$
and ${1\ov \ep^2} \sim \ln^2 \L$ ones.
As at the 1-loop level \rf{lji}, the logarithmic divergences are expected to cancel
at the end  between the bosonic and fermionic contributions.

At the same time, it is easy to see that $\ln^2 \L$  divergences should
cancel separately for bosons (and thus also separately  for   the fermions).
This follows from the basic  renormalization properties of the sigma model
in the case of the target-space metric corresponding to the Einstein space
$R_{\m\n} = k  G_{\m\n}$.
Indeed, let us recall few basic facts about 2d  sigma model renormalization
in dimensional
regularization (see, e.g.,  \ci{honer,frid,alv}).
Using subscript  $0$ to denote bare quantities  and $\mu$ for the  renormalization scale
we have for the partition function\foot{Since we will be expanding near a
classical solution, we will not
need to worry about field renormalization;  the  parameters of our background
cannot get renormalized.  As was  already mentioned above,
for a homogeneous solution
there is also no reason to expect any change in the form of the
background due to quantum corrections to the effective action.}
\be \la{ze}
Z_0 ( G_0, \ep) = Z(G, \mu) \ ,\ \ \ \ \ \ \ \ \
\mu {\del Z\ov  \del \mu}   + \beta \cdot {\del Z\ov \del G} = 0  \
 .\ \ee
Here  $d= 2-2 \ep, \  { 1 \ov \ep} \sim    \ln \L \to  \infty $  and
\be \la{bar}
G_0 = \mu^{-2 \ep}   \big[  G +  { 1\ov  \ep} T_1 ( G) +
 { 1\ov  \ep^2} T_2 ( G) + ... \big]\ ,
\ee
so that from $ { d G_0 \ov d \ln \mu} =0$ we get
\be \la{gze}
\hat \beta = { d G \ov d \ln \mu} = 2 \ep  G + \beta\ , \ \ \ \ \ \ \ \ \ \ \ \
\beta = 2( 1 -  G \cdot {\del\ov   \del G}) T_1   \ , \ee
\be
( 1  - G \cdot {\del\ov   \del G} ) T_2 =   (1 - G \cdot {\del \ov  \del G} ) T_1
  \cdot {\del\ov   \del G} T_1 \ .
\ee
To the 2-loop order
\be \la{zme}
(T_1)_{\m\n} = \ha   { \a'} R_{\m\n} +  { 1 \ov 16 }\a'^2  R_{\m\a\b\g }R_{\n}^{\ \a\b\g }
+ ...\ ,  \ee
\be \la{ke}
(T_2)_{\m\n} ={ 1 \ov 16 } \a'^2  \big(  D^\a D_\a  R_{\m\n} - D^\a D_\n R_{\m\a}
- D^\a D_\m R_{\n\a} + D_\m D_\n R \big)  \ , \ee
\be\la{me} \b_{\m\n} =
\a'  R_{\m\n} + { 1 \ov 2 }\a'^2  R_{\m\a\b\g }R_{\n}^{\ \a\b\g } + ...\ .  \ee
In the case when  the metric is the  direct product of
 the $AdS_5$ and $S^5$ parts  the Ricci tensor is  covariantly constant   so that
 for each  factor  $T_2=0$, i.e. there are no
 ${1 \ov \ep^2 }\sim \ln^2\L$ divergences.\foot{The cancellation
of  $\ln^2$-divergences implies  also the cancellation of $\ln$-divergences
with transcendental coefficients like $\ln 2$ or Euler constant  $\g$.}

For the $S^{N}$ sigma model (with radius $a$ playing the role of
the running coupling constant) we have
$R_{\m\a\b\g } = {1\ov a^2} ( G_{\m\b} G_{\a\g} -  G_{\m\g} G_{\a\b}  ),
\  \ R_{\m\n} = {(N-1)\ov a^2}  G_{\m\n}$  so that
\be \la{tt}
(T_1)_{\m\n} = {N-1\ov 2\sql  }\big[1 +     {1  \ov 4\sql  }    + O({1  \ov (\sql)^2  })
 \big] G_{\m\n} \  ,  \ \ \ \ \ \ \ \ \sql = { a^2 \ov \a'} \ .
\ee
The corresponding 2-loop beta-function is of course the same
as for the $O(N+1)$ sigma model    \ci{pol,brez}, i.e. (for $\a'=1$)
we get  $\beta =  { d a^2 \ov d \ln \mu } =
(N-1) ( 1 + {1\ov a^2} ) + ...$.
For $AdS_N$ one needs to invert the sign of the  first term
($a^2 \to - a^2$).

The  coefficients
of the logarithmic divergences in the \sm effective action computed  in
 a particular background
should  be consistent   with these general results.
The divergent part of the  effective  action should  be  cancelled by the
cutoff dependent terms in the bare sigma model action. Evaluated on the background
\rf{rre} the latter  is given by (for the $S^5$ part of the bosonic action, $N=5$)
\bea
I_0&=&  { \sql \ov 4 \pi} \ \mu^{-2 \ep}
  ( 1 + { 2 \ov \sql\ \ep } + { 1 \ov (\sql)^2\ep  } + ... )
\int d \tau \int^{2 \pi}_0 d \s  \ ( m^2 - w^2 )\no \\
 &=&
- \k^2 \mu^{-2 \ep}  ( \sql + {2\ov \ep} + { 1 \ov 2 \sql\ \ep } + ... )\int d \tau   \ , \la{oy}
\eea
where we used that in  the   scaling
limit  $m^2 \to -w^2=\k^2  $  we get $m^2 - w^2 \to - 2\k^2$.\foot{
The 1-loop coefficient here agrees with the UV divergent
term coming from the bosonic part of the 1-loop effective action \rf{lji}.
Note that  $ \ha [\ln \det ( - \del^2 + M^2) ]_\infty =  { 1 \ov 4 \pi}
V_2 M^2 \ln { \Lambda \ov \mu}
= { 1 \ov 8 \pi \ep}
V_2 M^2 $.}

\subsection{\adss  sigma model fluctuation action }
As a preparation for the 2-loop computation of the effective action
let  us now consider the \adss   bosonic action in conformal gauge
 expanded near
the background \rf{rre},\rf{sci} to quartic  order in fluctuation fields.

We shall adopt the following parametrization  of the $AdS_5$ and $S^5$ parts of the
metric
\begin{eqnarray}
&&\label{metric}  ds^2 = (ds^2)_{AdS_5} + (ds^2)_{S_5} \ , \\
&&(ds^2)_{AdS_5} =
-\big(\frac{1+\frac{1}{4}z^2}{1-\frac{1}{4}z^2}\big)^2 dt^2
+\frac{dz^kdz_k}{(1-\frac{1}{4}z^2)^2} \ , \ \ \ \ \  \ \ \ \ \ k=1,2,3,4 \ ,
\la{ads}\eea
\bea
(ds^2)_{S_5}=\frac{dx^2+dy^2-(x dy-y dx)^2      }{1-x^2-y^2}
+(1-x^2-y^2)\left(d\psi^2 +\cos^2\psi\ d\phi_2^2 +\sin^2\psi\
d\phi_3^3\right) \ .
\la{ssm}
\end{eqnarray}
The somewhat  unusual form of the $S^5$ metric is chosen so
that to have a regular expansion near the  $S^3$ solution \rf{rre}.\foot{
The standard metric $(ds^2)_{S_5}= d \theta^2 + \cos^2 \theta\ d \phi_1^2
+ \sin^2 \theta\ (d \psi^2 +\cos^2\psi\ d\phi_2^2 +\sin^2\psi\
d\phi_3^3)$   is related to the above one
by the following coordinate transformation:
$x = \cos \theta \cos \phi_1, \ \ x = \cos \theta \sin \phi_1$.}
As discussed in section  1.1  above, we will be interested in  the special case of the  formal
analytic continuation \rf{rrr}
of this solution  with the parameters given by \rf{sci}, i.e.
\bea
&&t=0, \ \ \ \ \  z_k =0 \ , \ \ \ \ \ \
  ~~x=0~, \ \ \ \ \ \ ~~~y=0~, \no \\
&& \psi={\pi\ov 4 }
~,~~~~~~~~~~\phi_2=\k (\tau-\sigma)~, ~~~~~~~~
\phi_3=\k (\tau+\sigma)\ , ~~~~~~
\label{ro}
\eea
Expanding the bosonic part of the string action
 to quartic order in fluctuations near
 this   background
 \bea
&&t= \tilt~, \ \ \ \  ~~~z_k={\tilde z}_k~,~~~\ \ \ \ \
x={\tilde x}~, \ \ \ \ ~~~
y= {\tilde y}~,\ \ \ \ \
~~~~\psi={\pi\ov 4}+  {\tilde \psi}\ , \  \cr
&&
\phi_2=\k(\tau-i \sigma)+{\tilde \varphi}_2-{\tilde\varphi}_3\ , \ \ \ \ \ ~~~~
\phi_3=\k(\tau+i \sigma)+{\tilde\varphi}_2+{\tilde\varphi}_3 \ ,  \la{jio}
\eea
 we get for the quadratic, cubic and quartic terms  in the bosonic action
  \be
  I_B = \int  d\tau \int^{2\pi}_0 d \s\  {\cal L}_B =
 - {\sql }\ \k^2 \int d\tau  +  \int  d\tau \int^{2\pi}_0 d \s\
  ({\cal L}_2 + {\cal L}_3 + {\cal L}_4+ ...) \ , \la{actp}\ee
\bea
{\cal L}_2&=&-\frac{\sqrt{\lambda}}{4\pi}\Big[
-(\partial_\alpha {\tilt})^2+(\partial_\alpha \td z_k)^2
+(\partial_\alpha {\tilde x})^2
+(\partial_\alpha {\tilde y})^2
+2\k^2 {\tilde x}^2+2\k^2{\tilde y}^2
\cr
&&~~~~~~~~~
+(\partial_\alpha {\tilde \psi})^2
+(\partial_\alpha {\tilde \varphi}_2)^2
+(\partial_\alpha {\tilde \varphi}_3)^2+4\k\,{\tilde\psi}
(\partial_\tau{\tilde\varphi}_3+i\,\partial_\sigma{\tilde\varphi}_2)
~~\Big]\ ,
\label{L2}
\eea
\bea\la{yu}
{\cal L}_3=-\frac{\sqrt{\lambda}}{4\pi}\Big[~
2\k\,({\tilde x}^2+{\tilde y}^2)
\left(\partial_\tau{\tilde \varphi}_2 + i \partial_\sigma{\tilde \vp}_3\right)
-4{\tilde \psi}
\ \partial_\alpha {\tilde \varphi}_2
{\partial^\alpha {\tilde \varphi}_3} ~~\Big]\ ,
\eea
\bea
{\cal L}_4&=&\frac{\sqrt{\lambda}}{24\pi}
\Big[\ \ \ 3 (\td z_k)^2\left(-2 (\pa_\alpha\tilt)^2+ (\pa_\alpha \td z_n)^2\right)~
\no \\
&&~~~~~~~
-6 \left({\tilde x}^2+{\tilde y}^2\right)
\left((\pa_\alpha{\tilde\varphi}_2)^2+(\pa_\alpha{\tilde\varphi}_3)^2
+(\pa_\alpha{\tilde\psi})^2+4 \k \,
{\tilde\psi} \left(\pa_\tau{\tilde\varphi}_3+i \pa_\sigma{\tilde\varphi}_2
\right)\right)\no \\
&&~~~~~~~
+ 6 \left({\tilde x}\pa_\alpha{\tilde x}+{\tilde y}\pa_\alpha{\tilde
y}\right)^2
- 16  \k \,{\tilde\psi}^3 (\pa_\tau{\tilde\varphi}_3+i\pa_\sigma{\tilde\varphi}_2)
\Big] \ .
\label{L4}
\eea
Let us now make a few  remarks.

Since the background  values in  \rf{ro},\rf{jio}
depend on $\k$  only in combination with world-sheet coordinates,
we can  factorize the  $\k$-dependence
in the  Lagrangian ($\cL \to \k^2 \cL$)
by making the  rescaling
$$ \k \tau \to \tau,\ \ \ \ \ \ \ \    \ \ \k \s \to \s \ .  $$
This rescaling gives an equivalent theory assuming that  scale invariance
survives at the quantum level;
this is not the case   in the pure bosonic theory but should be so once fermions are added.

After the rescaling by $\k$ (and assuming the cutoff dependence cancels out at the end)
the  string action on $R_\tau \times (S^1)_\s$ will
depend on $\k$  through the upper
 limit of integration $2\pi \k$ over rescaled $\s$.
 In the limit $\k\to \infty$ we are  interested in
 we can then decompactify the spatial world-sheet dimension
 and thus use momentum representation with  continuous spatial
 components.


The 1-loop correction to the effective action that follows from \rf{L2}
can be easily seen
to be
 in agreement with the bosonic part of \rf{pi},\rf{lji}.
The quadratic part  of the fluctuation   action \rf{L2} can be diagonalized
by a (non-local) ``rotation'' of the three $S^3$ fields (see   \ci{art}).
This will bring in one  massive  and two massless  modes in the $(\td \psi,\td
  \vp_2,\td \vp_3)$ sector.   The
resulting  quadratic fluctuation part of the superstring
action will have 
the form of 2d Lorentz invariant collection of massive bosonic and
fermionic  fields, but  higher-order terms in fluctuations 
will no longer have 2d Lorentz invariance (which is ``spontaneously broken''
 by our choice of the
background). Expressed in terms of the ``rotated'' fields the interaction terms will have
non-local form. For that reason here we
 choose not to perform this diagonalization explicitly and
use non-diagonal propagator instead.

As was  already mentioned, in
  conformal gauge the bosonic contributions of $AdS_5$ and $S^5$ parts
factorize.
If we  formally set $\k=0$ in \rf{L2},\rf{yu},\rf{L4}, i.e. consider
the case of trivial background in all directions, then the  $AdS_5$ and $S^5$
contributions to the partition function will become similar.\foot{The fact that
in the $AdS_5$  part we have only quartic interaction while in the
 $S^5$ part we also have a cubic one is an artifact of a particular
  parametrization  and the choice of the expansion point
 used.}



In the  action \rf{actp} we assumed the
Minkowski world-sheet  signature $(-,+)$;  the action is not real  because of
our choice of the imaginary value of the winding parameter
$m$. The Euclidean action obtained by
continuing $\tau \to i \tau$  is also not real  but the  imaginary parts are
linear in
$\k$  and derivatives, so the partition function and the
effective action will be real.
We shall continue to Euclidean signature
 at the level of momentum-space integrals.

\subsection{Structure of 2-loop quantum corrections}

The   computation  we shall describe below may be 
viewed as a special case of  computation of 2-loop 
correction to a mass of a sigma model soliton. 
In general, the mass is determined   by a logarithm of the 
partition function  computed on a time interval with soliton boundary conditions 
\ci{deve}. In the present case of a homogeneous field configuration
it turns out that there is no distinction between the connected and simply connected
graphs, so we shall   consider the 2-loop correction to the 
 1-PI  effective action. Also, the homogeneous (``delocalized'') nature of the field 
 configuration implies there is no   non-trivial issue of separation of the 
 contribution of zero modes (cf. \ci{deve}).

\begin{figure}[ht]
\centerline{\includegraphics[scale=0.5]{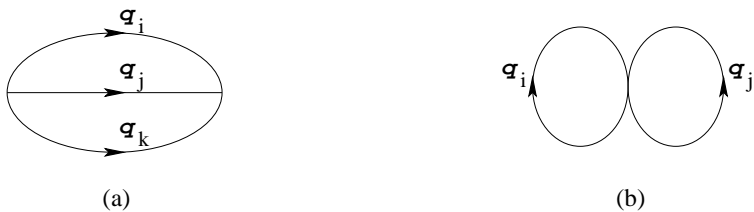}}
\caption{Two-loop contributions (momentum conservation  $q_i+q_j+q_k=0$
is assumed).
\label{topologies}}\nonumber
\end{figure}

The 2-loop  contributions to the effective action
in   a theory like \rf{actp}
with three-point  and four-point vertices is given by
the Feynman diagrams of the  two topologies shown in figure \ref{topologies}.
In general, the  lines  in these diagrams may be either bosons or fermions.
The 2-loop 1-PI effective action  is then  given by
\be\la{gag}
\G = V_2 \bG\ , \ \ \ \ \ \ \ \ \  \bG = \bG_1 + \bG_2 + ... \ , \ \ \ \ \ \ \ \
\bG_{\rm 2}=
\bG_{\rm cubic}+
\bG_{\rm quartic}+\delta \bG_{\rm measure} \ .
\ee
Here $V_2$ is the volume factor as in \rf{pi} (our background is homogeneous),
i.e.  $\bG$ stands for the effective Lagrangian.
$\delta \bG_{\rm measure}$  is the contribution
coming from the measure  counterterm \rf{mea}
expanded to quadratic order in fluctuations, i.e.  (after Wick rotation)
\be\la{quaa}
\delta \LL=-\ha \big[3\td z^2-2{\tilde x}^2
-2{\tilde y}^2- 4{\tilde \psi}^2 + O(\td \p^3) \Big]\int \frac{d^d q_j}{(2\pi)^d\mu^{d-2}}\ ,
\ee
where $ \int \frac{d^d q_j}{(2\pi)^d\mu^{d-2}}$ is the (correctly normalized) integral
representation of
%
%
$\delta^{(2)}(0)$.
 The insertion of this counterterm
into a 1-loop diagram  will cancel all  quadratic divergences in the
2-loop effective action.
 We will be using    the dimensional
regularization  with $\mu$ as a renormalization scale and
 $d=2-2\ep$
since this is an invariant regularization preserving the symmetries of the
sigma model.
Power divergences can be ignored  in dimensional regularization
but it is sometimes useful to track their cancellation against the measure
as a check of combinatorial factors.

\bigskip

To compute the 2-loop diagrams  we need to work out the  propagator.
The quadratic terms in \rf{L2}  contain off-diagonal mixings which can be readily
diagonalized as in  \ci{art}. However, we found it more convenient
%
%
to keep the propagator  off-diagonal.
 Ordering the fluctuation fields as follows
\be \la{phi}
\P_i= \{\td t; (\td z_1,\td z_2, \td z_3, \td z_4 ); (\td x, \td y);( \td \psi, \td \vp_2, \td \vp_3)\}
\ee
one finds  from \rf{L2}
\be
\Delta^{-1}(q)=\frac{2\pi}{\sqrt{\lambda}}\begin{pmatrix}
-\frac{1}{q^2} & 0&0&0&0&0\cr
0&\frac{1}{q^2}\,\id_4 &0&0&0&0\cr
0&0&\frac{1}{q^2+2}\,\id_2 &0&0&0\cr
0&0&0& \frac{1}{q^2+4} &
\frac{2q_1}{q^2(q^2+4)}&
\frac{-2i\,q_0}{q^2(q^2+4)}\cr
0&0&0& -\frac{2q_1}{q^2(q^2+4)}&
\frac{(q^2)^2-4 q_0^2}{(q^2)^2(q^2+4)}&
\frac{2i\,q_0q_1}{(q^2)^2(q^2+4)}\cr
0&0&0&\frac{2i\,q_0}{q^2(q^2+4)}&
\frac{2i\,q_0q_1}{(q^2)^2(q^2+4)}&
\frac{(q^2)^2+4 q_1^2}{(q^2)^2(k^2+4)}
\end{pmatrix}\la{pro}
\ee
Here $q_\a= (q_0,q_1)$ is 2-momentum. We have
rescaled the coordinates by $\k$ (with $\k \to \infty$)
and will assume that momenta take  continuous values.
Continuation to Euclidean signature is done by $q_0 \to  - i q_0$.
This  eliminates $i$-factors from the propagator.
The 1-loop effective action is then
$\G_1 =  \ha \Tr \ln \Delta$ and agrees with \rf{pi},\rf{lji}.\foot{We
may formally ignore the ``ghost'' nature of the $\td t$ fluctuation and then
the 1-loop contribution of two massless ``longitudinal'' modes
is cancelled by  the  conformal gauge ghost contribution to the
 partition function. The ``ghost'' sign of the time  direction is irrelevant also
 for the higher-loop corrections:   since time direction enters the action only
 quadratically, it can be integrated out once and for all
 (e.g., with $t\to i t$ prescription to make Euclidean path integral convergent)
 and that does not lead to any sign changes  compared to the case
 when $t$ would have ``physical'' sign.}

\bigskip

Defining the cubic vertex as $V_{ijk}(q_i,q_j,q_k)=\frac{\pa^3
 \LL}{\pa\Phi_i\pa\Phi_j\pa\Phi_k}\big|_{\Phi=0}
$, i.e. writing the (Euclidean) fluctuation Lagrangian  corresponding to  \rf{actp} as
\be
\LL = \ha \Phi_i  \Delta_{ij} \Phi_j
+  { 1 \ov 3!} V_{ijk}\Phi_i  \Phi_j  \Phi_k
+  { 1 \ov 4!} V_{ijkl}\Phi_i  \Phi_j  \Phi_k\Phi_l + ...
\la{vvvv}
 \ ,
\ee
we can compute the contribution of the graphs with topology (a) in figure
\ref{topologies}  as
\bea
\bG_{\rm cubic}&=&c_3
\int\frac{d^d q_id^d q_j}{(2\pi)^{2d}\mu^{2d-4}}\,
V_{ijk}V_{i'j'k'}\Delta^{-1}_{ii'}\Delta^{-1}_{jj'}\Delta^{-1}_{kk'}\cr
&=&c_3
\frac{4\pi}{\sqrt{\lambda}}
\int\frac{d^d q_id^d q_j}{(2\pi)^{2d}\mu^{2d-4}}(\I_{1}+\I_{2}+\I_{3}+\I_{N})
 \la{ttt}\eea
where\foot{Here we assume Euclidean continuation,
i.e. \ $e^{-\G} = \int [d\Phi] \ e^{-S}, \ \ S= \int d^2 \s \LL $.}
$$c_3 = - { 1 \ov 12} $$
is the combinatorial factor of the diagram and we have solved the vertex momentum
conservation constraint by setting $q_{k}=-(q_i+q_j)$.
We assume that continuation to $d$ dimensions
is done at the level of the momentum integrals, and $\mu$-factors are
introduced to balance the dimensions.
The overall factor of $\k^2$ is included
 in the volume $V_2$ in \rf{gag}
as in \rf{pi},\rf{ken}.

There are many
equivalent expressions for
the integrands $\I_1$, $\I_2$, $\I_3$ and $\I_N$; the
one which exposes both
the UV and IR convergence properties of the loop integrals is:\foot{Here $q_i$ and $q_j$
denote two momenta without any summation over $i,j$  and $q_{k}=-(q_i+q_j)$.}
\bea
&&\I_{1}=3
\frac{4}{q_i^2+4}\ , \la{h} \\
&&\I_{2}=3 \,\left[-\frac{2}{q_i^2q_j^2}
+\frac{4}{q_i^2(q_j^2+4)}
-\frac{4}{\left(q_i^2+2\right)(q_j^2+2)}
-\frac{14+\frac{8}{d}}{\left(q_i^2+4\right)(q_j^2+4)}\right]\ ,
\la{hh} \\
&&
\I_{3}=3\,\bigg[\frac{8}{q_i^2q_j^2}+\frac{16}{
\left(q_{i}^2+2\right)
\left(q_{j}^2+2\right) }
-
\frac{8}{
\left(q_{i}^2+4\right)
\left(q_{j}^2+4\right)}
\bigg]\frac{1}{q_{k}^2+4}\ ,
\la{hhh} \\
&&\I_{N}=3 \,\Bigg[(q_{i1}q_{j0}+q_{i0}q_{j1})^2
\frac{8(q_i^2+q_j^2-q_k^2)^2}
{(q_i^2)^2 \left(q_i^2+4\right)
 (q_j^2)^2\left(q_j^2+4\right)    \left(q_k^2+4\right)}
\cr
&&~~~~~~~~~~~~
-(q_{i0}^2-q_{i1}^2)^2
\frac{16}{(q_i^2)^2 \left(q_i^2+4\right)
\left(q_j^2+2\right)\left(q_k^2+2\right)}
\la{inti}\\
&&~~~~~~~~~~~~
-(q_{i0}q_{j0}-q_{i1}q_{j1})(q_{i0}q_{k0}-q_{i1}q_{k1})
\frac{16\left[(q_i^2)^2-(q_j^2-q_k^2)^2\right]}{
(q_i^2)^2 \left(q_i^2+4\right)
 q_j^2\left(q_j^2+4\right)q_k^2\left(q_k^2+4\right)}
\Bigg]
\nonumber
\eea
In \rf{hh} $d=2-2\ep$.\foot{The factor $1\ov d$ in \rf{hh}
 came from a reduction of a tensor integral to a scalar integral due to
symmetric integration:
$
\int \frac{d^d q_id^d q_j}{(2\pi)^{2d}}
   {(q_i\cdot q_j)^2\over (q_i^2+4 )(q_j^2+4 )}
    = d^{-1}  \int \frac{d^d q_id^d q_j}{(2\pi)^{2d}}
     { q_i^2 q_j^2\over (q_i^2+4 )(q_j^2+4)} $.
In general, $
\int\frac{d^d q}{(2\pi)^{d}}  {q^\a q^\b\over (q^2+4)^n }
=   d^{-1}      \int \frac{d^d q}{(2\pi)^{d}}   {  \eta^{\a\b} q^2\over (q^2+4)^n} $.}
We also continued to Euclidean space by replacing $q_{j0}  \to - i q_{j0}$, so that
in the above expressions  $q^2_j = q^2_{j0} + q^2_{j1}$.

$\I_{1}$ and $\I_{2}$ give rise to UV-divergent integrals;
the integral of $\I_{1}$ contains  power-like divergences and  the integral of
$\I_{2}$ --
logarithmic divergences.
The first two terms in $\I_{2}$ and the first term in
$\I_{3}$ give rise to IR-divergent integrals.
In addition to the  dimensional regularization
for the UV divergences  we shall introduce a small
mass parameter $m_0$ to regularize the IR divergences.\foot{We will not use
regulators in finite integrals.}

The subscript $N$ on $\I_N$ is used to indicate  that this integrand does
not look  2d Lorentz invariant. However, the integral of $\I_N$
(which is  UV and IR finite)
can be expressed in terms of
  Lorentz-invariant integrals.
While  the original sigma model action (the string action in conformal gauge)
is 2d  Lorentz-invariant, this symmetry is spontaneously broken by a choice of
the background in \rf{ro},\rf{jio}, i.e. (cf. \rf{ssm})
\be \la{neq}
\cos \psi d\p_2 = N_\a d\s^\a,\ \ \ \ \
 \ \sin \psi d \p_3 = N^*_\a d\s^\a,\ \ \ \ \ \ \
 \ N_\a= {\k \ov \sqrt 2 }(1,-i),\ \ \  \ N^*_\a= {\k \ov \sqrt 2 }(1,i).   \ee
The 2-loop effective action then depends on the background through
the mass  terms (proportional to $N^*_\a N^\a =- \k^2$, etc.)
and also through the explicit factors of $N_\a $ and $N^*_\a$
in the denominators  of momentum  integrals. Indeed, $\I_N$ in \rf{inti}
  is proportional to 4 factors of
these vectors.
Since the rest of the momentum integrands are Lorentz-covariant,
they can be reduced to  products of contractions  between $N_\a $ and $N^*_\a$
factors and scalar Lorentz-invariant momentum integrals. We shall illustrate
how that happens below.
As a result,  the corresponding term in $\G_2$ will
contain 4 factors of  first derivatives of the
background fields, i.e. will be proportional to
$ \del^\a \p_2 \del_\a \p_2 \del^\b \p_3 \del_\b \p_3 + ...$
with coefficients that are given by Lorentz-invariant momentum
 integrals.\foot{Let us note that the use  of dimensional regularization in a situation with
 Lorentz  invariance spontaneously broken by either the  background or by gauge choice is not
 uncommon (cf., e.g.,  discussions of YM theory in lightcone gauge \ci{BDKlc}).}

\bigskip

Similarly, for the contribution of the  quartic vertex in \rf{vvvv}
$V_{ijkl}=\frac{\pa^4 \LL}{\pa\Phi_i\pa\Phi_j\pa\Phi_k\pa\Phi_l}
\big|_{\Phi=0}$
to the diagram (b) in
figure \ref{topologies} we find
\be \la{qui}
\bG_{\rm quartic}=c_4
 \,\int\frac{d^d q_i d^d q_j}{(2\pi)^{2d}\mu^{2d-4}}\,
V_{ijkl}\Delta^{-1}_{ij}\Delta^{-1}_{kl} =c_4
 \,\frac{4\pi}{\sqrt{\lambda}}
\int\frac{d^d q_i d^d q_j}{(2\pi)^{2d}\mu^{2d-4}}\,(\J_{1}+\J_{2}) \ ,
\ee
where $$c_4= {1 \ov 8}$$
 is the  combinatorial factor.
Despite the relatively  complicated-looking
quartic   Lagrangian (\ref{L4}) the integrands
$\J_1$ and $\J_{2}$ are very simple:
\bea
&&\J_{1}=\frac{24}{q_i^2}-\frac{8}{q_i^2+2} \ , \la{iii}
\\
&&\J_{2}=-\frac{8}{(q_i^2+2)(q_j^2+2)}
-\frac{32}{(q_i^2+4)(q_j^2+4)} \ .  \la{iiii}
\eea
Both $\J_{1}$ and $\J_{2}$ lead to UV-divergent integrals --
power-like and logarithmic, respectively.

The contribution of the measure counterterm \rf{quaa} is
\be\la{mik}
\delta \bG_{\rm measure}=-\frac{4\pi}{\sqrt{\lambda}}\,
\int\frac{d^dq_id^dq_j}{(2\pi)^{2d}\mu^{2d-4}}
\,\Big(\frac{3}{q_i^2}
-\frac{1}{q_i^2+2}
-\frac{1}{q_i^2+4}\Big) \ .
\ee
It is not hard to check that it cancels all power-like  divergences in the
2-loop integrals in \rf{ttt} and \rf{qui}.
In particular, it cancels the contribution of the $\J_1$ integral in  \rf{iii}.

\bigskip

Let us note that  if we  formally  consider the theory
  \rf{actp} defined on
  $R \times R$   and set $\k=0$   then the  corresponding 2-loop
  effective action will be  given by \rf{gag} with  \rf{ttt} containing
  only ``massless'' limit   $  12 \ov q^2_i$  of $\I_1$ in
  \rf{h}
 and with \rf{qui} containing  only the
  ``massless'' limit  $  16 \ov q^2_i$ of  $\J_1$
 in \rf{iii}.  Their sum is then cancelled by the ``massless'' limit
 of the measure  contribution \rf{mik} (with the integrand $   1 \ov q^2_i$).
 Thus $\G_2 (\k\to 0) \to 0$.

\subsection{Evaluation of 2-loop momentum integrals}

Combining the above 2-loop contributions we get for \rf{gag}
\bea
\bG_{\rm 2}&=&
\bG_{\rm cubic}+
\bG_{\rm quartic}+
\delta \bG_{\rm measure}\cr
&=&\,\frac{4\pi}{\sqrt{\lambda}}\,
\int \frac{d^dq_id^dq_j}{(2\pi)^{2d}\mu^{2d-4}}
\left[\big(-\frac{1}{12}\I_{2}+\frac{1}{8}\J_{2}\big)
-\frac{1}{12}(\I_{3}+\I_{N})\right] \ . \la{dii}
\eea
Here the contribution of the
first parenthesis contains all UV divergences.
It turns out  that the
contributions of states with mass-squared equal to $2$ cancel
between the  topologies (a) and (b). Then we get
($d=2-2\epsilon$)
\be\la{hjl}
-\frac{1}{12}\I_{2}+\frac{1}{8}\J_{2}=
\frac{\epsilon}{1-\epsilon}\frac{1}{(q_i^2+4)(q_j^2+4)}+
\frac{1}{2} (\frac{1}{q_i^2}-\frac{1}{q_i^2+4})
(\frac{1}{q_j^2}-\frac{1}{q_j^2+4}) \ .
\ee
The contribution of the second  term in \rf{hjl}
 is  UV-finite but IR-divergent.
 As
was  mentioned above, we shall
regularize this IR divergence  by introducing a small
mass $m_0$. Using the standard  integral
\bea
I(M^2)\equiv
\mu^{2\epsilon}\int\frac{d^d q}{(2\pi)^d}\frac{1}{q^2+M^2}
=\frac{1}{(4\pi)^{d/2}}\frac{\pi}{\Gamma(2-\epsilon)\,\sin(\pi\epsilon)}
\left(\frac{\mu^2}{M^2}\right)^\epsilon\cr
 \approx { 1 \ov 4 \pi} \big[ { 1 \ov \ep} +
 1 - \g  + \ln { 4 \pi \mu^2 \ov M^2}  + O(\ep) \big]
 \ , \la{oop}
\eea
we then   find\foot{Here $\g= - \Psi(1)=0.5772...$  is the Euler constant.
  Let us also recall that we have rescaled
the world-sheet variables  by $\k$. If we did not do this but still formally
decompactified  the spatial direction of the world sheet we would get
 the first term here as
 $$\bG_{\rm 2}=
\frac{\k^2}{4\pi\sql }\big[\frac{1}{\epsilon}+3-2\gamma+
2\ln\frac{\pi\mu^2}{\k^2}+ \frac{1}{2}
\ln^2 (\frac{m_0^2}{4\k^2 })\big] +  {\rm finite}\ .$$}
\be\la{res}
\bG_{\rm 2}=
\frac{1}{4\pi\sql }\left[\frac{1}{\epsilon}+3-2\gamma+
2\ln (\pi\mu^2) +\frac{1}{2}
\ln^2(\frac{m_0^2}{4}) \right]
-\frac{4 \pi }{12\sql }
\int \frac{d^2q_id^2q_j}{
(2\pi)^{4}}\left(\I_3+\I_N\right) \ .
\ee
As expected for a symmetric-space sigma model, the double-pole $\frac{1}{\epsilon^2}$
 UV divergences
cancelled out (cf. \rf{ke},\rf{oy}).
The effective action is found  by multiplication of this
expression by $V_2  =2 \pi \k^2 \bar T  $ as in \rf{ken}.


\bigskip

Next, let us  compute the  integral of $\I_3$ in \rf{hhh},\rf{dii}, writing it as
\be\la{ijk}
I_3 =  \int\frac{d^2q_id^2q_j}{(2\pi)^{4}} \ (\I_{3,1} + \I_{3,2} + \I_{3,3})
= I_{3,1} + I_{3,2} + I_{3,3} \ . \ee
The  integral of the first term
\be \la{uul} I_{3,1}=\int\frac{d^2q_id^2q_j}{(2\pi)^{4}}
 {24 \ov q^2_i q^2_j ((q_i + q_j)^2 +4)} \ee
  with two massless propagators
 is IR divergent and we need to regularize it  by  $m_0 \to 0$.
 This  leads to an integral which is a special case  of the following
  integral
 with
 3 massive propagators with at
least two equal masses\foot{We may solve the momentum conservation condition as
$q_k=-(q_i+q_j)$ or as $q_j= - (q_k + q_i)$;  the final result is the same.}
\be\la{ip}
I(M,M')
=\int\frac{d^2q_id^2 q_j}{(2\pi)^4}\frac{1}{(q_i^2+M^2)(q_j^2+M'^2)[(q_i+q_j)^2+M'^2]}
\ee
The calculation  of this integral is standard:
we Feynman-parametrize the propagators with equal masses and do the
integral over $q_j$ with the result:
\be\la{uuv}
I(M,M')=\frac{1}{4\pi}
\int_0^1dx \int\frac{d^2q_i}{(2\pi)^2}\frac{1}{(q_i^2+M^2)[x(1-x)q_i^2+M'^2]}
\ee
There is no need of  Feynman parametrization for the second momentum
integral; computing it directly leads to
\be
I(M,M')=\frac{1}{(4\pi)^2}
\int_0^1dx\ \frac{\ln\frac{M^2}{M'^2}+\ln [x(1-x)]}{x(1-x)M^2-M'^2} \ .
\la{uui}
\ee
For generic values of $M$ and $M'$ the remaining integral
  leads to a hypergeometric function.
However,  \rf{uul} corresponds to $M=2, \ M'=m_0 \to 0$.
Expanding \rf{uui} in $M'=m_0\to 0 $ we get for \rf{uul}
\be
\la{yyy}
(I_{3,1})_{m_0 \to 0}  =  { 6 \ov (4 \pi)^2 } \bigg[
{13\ov 3} \pi^2  + \ln^2  ( {m_0^2 \ov 4} ) \bigg] \ .
\ee
Multiplying this  by the
$ - { 1 \ov 12} {4 \pi\ov \sql}$ factor in \rf{res}
we conclude that the IR divergence from $I_{3,1}$
cancels the one in \rf{res}, so that the bosonic part of the
effective action is IR finite.\foot{This is of course what
one should have expected since we are computing  a physical quantity:
the value of the  (global symmetry invariant)
effective action on a classical solution, cf. \ci{dav,van}.}

 For the second term  $\I_{3,2}$ we need \rf{ip} with
  $M^2=4$ and $M'^2=2$, with \rf{uui}
then giving
 \bea &&
I_{3,2}=  \int\frac{d^2q_id^2 q_j}{(2\pi)^4}\frac{48}{
\left(q_{i}^2+4\right) \left(q_{j}^2+2\right)
\left( (q_i + q_j)^2+2\right)}\cr
&&\ \ \ \ \ \ = \frac{24}{(4\pi)^2}
\int_0^1dx\frac{\ln[2 x(1-x)]}{2 x(1-x)-  1 }
=\frac{48}{\,(4\pi)^2}  \KK \ .  \la{ka} \eea
Here $\KK$  is the Catalan's constant,
 \be \la{cata}
 \KK \equiv  \sum_{k=0}^\infty\frac{(-1)^k}{(2k+1)^2}=
 {1 \ov 16} \big[\Psi'(\textstyle{\frac{1}{4}})-\Psi'(\textstyle{\frac{3}{4}})\big]
=  0.915966...    \ ,
 \ee
where
 \be \la{psu}
\Psi'(z) = \psi_1(z) \equiv {d^{2} \ov dz^{2} }\ln \Gamma(z)
%
\  \ee
is the trigamma 
function.\foot{It admits the following series representation 
 $
  \psi_1(z) = \sum_{n = 0}^{\infty}\frac{1}{(z + n)^2}$
  and also satisfies 
  a reflection formula
 $   \psi_1(1 - z) + \psi_1(z) = \pi^2\csc^2(\pi z)$.
 Note also that    $\psi_1\left(\frac{1}{4}\right) = \pi^2 + 8\KK$.}
For the third
term in $\I_3$ in \rf{hhh},\rf{dii} we need \rf{ip} with  $M^2=M'^2=4$ so that
\bea &&
I_{3,3}= -
\int\frac{d^2q_id^2
q_j}{(2\pi)^4}\frac{24}{ \left(q_{i}^2+4\right)
\left(q_{j}^2+4\right) \left(  (q_i + q_j)^2+4\right)}\cr
&& \ \ \ \ \ \ \ =\
-\frac{6}{(4\pi)^2} \int_0^1dx\frac{\ln [x(1-x)]}{x(1-x)-1}
=-\frac{24}{\,(4\pi)^2} \  \td \KK    \ , \la{io}  \eea
where
\be \la{ce}
\td \KK  \equiv {1 \ov 72} \big[\Psi'(\textstyle{\frac{1}{6}})+\Psi'(\textstyle{\frac{1}{3}})
-\Psi' (\textstyle{\frac{2}{3}})-\Psi'(\textstyle{\frac{5}{6}})\big]
= 0.585976...  \ . \ee
Let us note also an alternative representation
for $\td \KK$   similar to the one  for $\KK$ in \rf{cata}
which follows from the series representation for $\Psi'(z)$ \foot{We 
thank M. Staudacher for  mentioning  this representation   to us   and  for 
  emphasizing that
 $\KK$ and $\td \KK$   have the same ``transcendentality''  (cf. \ci{rejs}).}
\be\la{lepa}
\td \KK = { 1 \ov 2}  \big[  \sum_{k=0}^\infty\frac{(-1)^k}{(3k+1)^2}    +
 \sum_{k=0}^\infty\frac{(-1)^k}{(3k+2)^2}  \big]  \ .
\ee
The calculation of the integral of  $\I_N$ in \rf{inti} is
described in Appendix E.

Combining the partial results \rf{yyy},\rf{ka},\rf{io} and
\rf{hp}  we  find that the terms proportional to 
$\td \KK$ {\it cancel out}  in the sum of the integrals of $\I_3$ and $\I_N$ 
in \rf{dii}
 and thus the final  expression 
for the bosonic contribution
\rf{res} to the 2-loop effective  Lagrangian is\foot{The
rational term  in  the finite part of \rf{res} also cancells out between
the $I_3$ and $I_N$ contributions.}
\bea
\bG_{\rm 2B}=
\frac{1}{4\pi\sql }\bigg(
 [ \frac{1}{\epsilon}+3-2\gamma+
2\ln (\pi\mu^2)]
 - { 2}  \KK
\bigg)    \ . \la{rest}
\eea
The  divergent part here  is consistent  with the general  form of the 
counterterm in \rf{zme},\rf{tt}: the divergence 
cancels in the combination of \rf{rest}  with the bare classical action 
in \rf{oy}.\foot{The RG equation in \rf{ze} is verified  by noting that the  
coefficient of $\ln \mu$ term in \rf{rest} is twice compared to 
the one in the 1-loop result (cf. \rf{oop}).}

The fermionic contribution is expected  to cancel the divergent part
 and the associated finite terms, i.e. the  square  bracket
in \rf{rest}.

The non-trivial
finite bosonic contribution to the 2-loop string coefficient $a_2$
of $\ln S$ in \rf{sca},\rf{uuu}  is  then   proportional  to
$ \KK$: it is
found as in \rf{ken},\rf{en}
by multiplying \rf{rest} by $2 \pi \k \approx 2 \ln S$
and changing the overall sign according to    \rf{cha},\rf{idi}. This gives
\be \la{trans}
a_{2B} =  {1 \ov \pi}  \KK   
  \approx  0.29156\ . \ee
  Surprisingly, this   matches the numerical value in \rf{kkkk}
  up to the  sign. 
However, we are still to include the  contribution of the
2-loop graphs involving  fermions  and  it indeed appears to 
reverse the sign of the total value of   $a_2$. 

\renewcommand{\theequation}{3.\arabic{equation}}
 \setcounter{equation}{0}

\section{Fermionic   contribution to the 2-loop effective action}

Let  us now  turn to the contribution to the 2-loop effective action  coming from
diagrams containing  fermionic propagators.
The relevant terms in the \adss Lagrangian
expanded near the background \rf{ro}
can be symbolically written as
\be
\cL_F= \ha \theta K \theta    +
(\theta  M_1  \theta)   Y_1 \Phi   +
 (\theta M_2 \theta)   \Phi Y_2
 \Phi  +
  (\theta M_3 \theta)( \theta M_4 \theta) \ .
  \la{hio}
  \ee
Here   $\Phi$  stands   for the bosonic fluctuation fields \rf{phi}
and $K, M_n,Y_k$  are  combinations of Dirac matrices, numerical
tensors  and  world sheet derivatives
of the form $ A + B^\a \del_\a$. Their explicit form follows directly
from  the relations given in Appendix A but are
rather lengthy so we will not give it explicitly here.

As was already mentioned in the Introduction,
because the fermionic kinetic term is only linear in derivative while
the interaction vertices   contain  up to two derivatives,
 the GS string  theory is
formally of non-renormalizable type;  this  will manifest itself in the presence of
higher  power divergences.

Assuming the theory is actually finite,   all  of  power divergences  are
expected to be cancelled by the contributions of  the path integral measure and
$\k$-symmetry ghosts  (see Appendix C for a discussion of this in the flat space case).
Alternatively,   one may choose to use  dimensional regularization in which
all power divergences are automatically set to zero.
Then the remaining  $\ln^2 \Lambda\sim  {1 \ov \epsilon^2}$ divergences
 should cancel
separately in the fermionic sector  while the  $ \ln \Lambda
 \sim{  1 \ov \epsilon}$ contributions
should cancel against the bosonic  divergence in \rf{rest}.

There are several  potential  ambiguities  in how one
deals with divergent integrals. Since the GS   action contains a
WZ type term with
 $\ep^{\a\b}$ tensor,
 this  creates a potential problem with direct application of dimensional
 regularization.\foot{Let us note also  that the
parameters   of the $\k$-symmetry transformations are 2d self-dual vectors.}
 We shall assume that the dimensional regularization is applied only to scalar integrals
 at the last stage (after  all power-divergent parts of the momentum integrands are separated),
 i.e.  that all tensor algebra  is done in $d=2$;  in particular, we shall
 assume that $\ep^{\a\b}$ is not continued away from $d=2$.\foot{This is somewhat
 different from the
   case of the bosonic sigma model with an antisymmetric tensor coupling
   \ci{us,et} where one could assume  that
 $\ep^{\a\b} \ep^{\g \d} =  f(d) ( \eta^{\a\g} \eta^{\b\d} - \eta^{\a\d} \eta^{\b\g})$
 where $f(d) = 1 + a ( d-2) + ...$, and then show that a
 regularization scheme ambiguity  related to the choice of the coefficient
  $a$ can
 be absorbed into a redefinition of the sigma model coupling parameters.}

Our  assumption will be that such a restricted dimensional regularization
 prescription  is consistent with the basic
$\k$-symmetry of the theory at the quantum  level.  This
 is by no means obvious
and a problem with $\k$-symmetry  gauge dependence of the
2-loop
 result that we will encounter below
appears to be  an indication  of a problem with
this prescription.\foot{The standard proof of
gauge-independence of on-shell  effective action  assumes that  gauge
symmetry in question is preserved  at the quantum level,
i.e. implicitly assumes the existence
of an invariant regularization (but the  power counting
 renormalizability of the theory is
of course not required).}

One natural choice  of the $\k$-symmetry  gauge  (used
 at the one loop  order in \ci{dgt,ft1}) is $\theta^1=\theta^2$.
 This gauge is possible in type IIB  string action where both Majorana-Weyl  fermions
 in the GS   action have the same
 chirality.
One of its  advantages is preservation of  global bosonic symmetries of the action.
More generally, we may consider the  gauge $\theta^1=k \theta^2$
where $k$ is a real parameter (see Appendix B).
Cancellation   of $k$-dependence in the resulting effective action, i.e. its gauge-choice
independence, would be a  check of consistency
 of our  computation procedure (in particular, of the regularization we use).

\bigskip

Let us  first comment on   the  structure of the fermionic 2-loop contributions
in the  simpler   case of $k=1$ gauge.
The   quadratic part of the gauge-fixed action
follows from \rf{yy},\rf{yyo}  and is  given by
\be\la{qua}
{\cal L}_{F2} =\sqrt{2}\k{\bar\theta} \left[
\Gamma_8(\pa_\sigma-i\pa_\tau)  -
\Gamma_9(\pa_\sigma+i\pa_\tau)  \right] \theta +2i\k^2{\bar\theta}
\Gamma_*\Gamma_8\Gamma_9\theta \equiv \ha \theta^T K \theta  \ .
\ee
This leads to the propagator (where we again rescaled the
momentum  by $\k$)
\be\la{prop}
 K^{-1} (q)=\frac{1}{4\sqrt{2}(q^2+1) }\left[\Gamma_8(q_0+iq_1)
+\Gamma_9(q_0-iq_1)-i\sqrt{2} \Gamma_*\Gamma_8\Gamma_9\right] \C \ .
\ee
As a result, all fermionic modes have mass   equal to  1,
while the bosonic modes in \rf{pro}   had masses equal to 0, $\sqrt2$ and $2$
 (cf. the corresponding 1-loop expression in \rf{lji}).

 There are 3 different types of  2-loop diagrams   involving the fermions (see \rf{hio}):

(i)  diagram   in Figure 1 (a) with  two  fermionic and one bosonic propagators
(we shall call it ``FFB'' since it originates from the Yukawa interaction in \rf{hio});

(ii) diagram   in Figure 1 (b) with one bosonic and one   fermionic propagators
(originating from the  ``FFBB'' interaction);

(iii) diagram   in Figure 1 (b) with two    fermionic propagators
(coming from   ``FFFF'' vertex).

The most non-trivial contribution
with the integrand containing  {\it two  }
 fermionic and {\it one} bosonic  propagator
may come only   from the  FFB diagram.
Thus on general grounds we may expect that the finite
part of the fermionic contribution
which should supplement  the finite bosonic contributions in
\rf{ka} and \rf{io}
 should be
given by a combination of two possible  finite integrals
of the general form \rf{ip}:\foot{The third possible integral
$I(0,1)=  \int \frac{d^2 q_i d^2 q_j
}{(2\pi)^{4}}\frac{1}{(q_i^2+1)(q_j^2+1) (q_i+q_j)^2}$
is IR divergent (cf. \rf{yyy})   and  does not give a non-trivial transcendental contribution
to the finite part. It does not actually appear in the result of the computation.}
\bea
&&I(\sqrt 2 ,1)=  \int \frac{d^2 q_i d^2 q_j
}{(2\pi)^{4}}\frac{1}{(q_i^2+1)(q_j^2+1) [(q_i+q_j)^2 + 2 ]}=
\frac{1}{8 \pi^2} \KK   \ , \la{jpy} \\
&&I( 2 ,1)=  \int \frac{d^2 q_i d^2 q_j
}{(2\pi)^{4}}\frac{1}{(q_i^2+1)(q_j^2+1) [(q_i+q_j)^2 + 4]}
= \frac{\ln 2}{8\pi^2}    \ , \la{df} \eea
where in computing the integrals we used \rf{uui}
and $\KK$  is again the  Catalan's constant as in \rf{cata}.

It turns out that  only $I( \sqrt 2 ,1)$ in \rf{jpy}
appears as a result of the actual computation of the
FFB graph.
This leads to the  conclusion
 that the  finite  fermionic contribution
  alters the coefficient of the $\KK$-term  in
\rf{rest},\rf{trans}.
Assuming   all other possible finite contributions like $\ln 2$
which accompany logarithmic divergences (as in the square  brackets in \rf{rest})
 should  cancel out,
 we are then led to
the following  final answer for the coefficient   $a_2$ in \rf{uuu} (cf.
\rf{rest},\rf{trans})
\be \la{ns}
a_2=a_{2B} + a_{2F}  = {1 \ov \pi}  (1 + c_F)  \KK  
  \ , \ee
where  the coefficient $c_F$  of the fermionic  contribution remains to be determined.
The  result for $c_F$ in the $k=1$ gauge (and also in the light-cone type gauge)
 appears to be  $c_F=-2$ (see below).

\bigskip\bigskip\bigskip

Let us now  turn to   some  technical details of the actual computation
of the fermionic graphs   we have done.
Since the fermions are Majorana (we choose them to be real), the vertices
 in fermionic bilinears
in the action should be
antisymmetrized, i.e. $M_k$ in \rf{hio} should stand for $\ha (M_k - M_k^T)$.\foot{The antisymmetrization
should apply also to derivatives in $M_k$  (in $Y_2$ one should symmetrize them).} Then the
2-loop contributions to the 1-PI Euclidean effective action $\Gamma= -[ \ln Z ]_{1-PI}$
 coming from \rf{hio}
 are given symbolically  by:\foot{If the Minkowski space action  is
 $S= \ha \Phi  \Delta \Phi + \ha  \theta K \theta + ...$   then
 $e^{iS}  = \exp [ - \ha \Phi  (i\Delta^{-1})^{-1} \Phi - \ha  \theta (i K^{-1})^{-1}
  \theta +...]$.}

FFB: \ \     $ - i^2 \times   i^3 \times   \Tr[  M_1 K^{-1}(p) M_1  K^{-1}(-q)] Y_1  Y_1 \Delta^{-1}$
%
%

FFBB:  \ \    $    i \times   i^2 \times  \Tr[  M_2 K^{-1} ]  Y_2 \Delta^{-1}             $

FFFF: \  \
$ - 4 i \times i^2 \times  \big(  \Tr[ M_3  K^{-1}] \Tr[  M_4  K^{-1}]
                         - 2  \Tr[M_3  K^{-1}(p) M_4 K^{-1}(-q)]\big)$

The total number of fermionic 2-loop Feynman graphs one needs to evaluate is
 around few hundred.
With the help of a {\tt Mathematica}-based computer program
we computed the resulting integrands in the
fermionic contributions
to the 2-loop effective action represented
in the form of the double momentum integrals  as in \rf{ttt},\rf{qui}.
We found that in the $\theta^1= k \theta^2$ gauge
the integrand   depends
on the gauge parameter $k$ through the combination
\be \xi= (k - k^{-1})^2    \ee
and, unfortunately,  this dependence does not cancel  automatically.
We have re-arranged the integrands so that to extract power divergences
(using transformations of the type 
$ { p^2 \ov p^2 + m^2 } = 1 - {m^2 \ov p^2 + m^2 }$); the latter
   were then  set to zero by switching on
  dimensional regularization.
We also used the expressions for momentum integrals from Appendix E.2.
As a result, we found that the $\ln^2 \Lambda \sim { 1\ov  \ep^2}   $
plus $\ln \Lambda \sim { 1\ov  \ep}   $    UV divergent part
  in the 2-loop effective action
is coming from   (cf. \rf{gag},\rf{dii})\foot{The resulting effective action computed directly
in $d=2$  contains no IR divergences.}
\bea
&& ~~~~~~~~~~~~\bar  \G_{2F} =  { 2 \pi \ov \sql}   X   \ , \la{gem} \\
&& X_\infty = \Big(-8[1,2]- (4 -6\xi) [1,4]+(4-2\xi)[1,1]\Big)
+\Big(8[1,2]+4[1,4]\Big)\cr
&&+ \ 
%
\frac{2}{3}
(4+ \xi )(-36+10+0+40)[1,1]
=  6 \xi [1,4]   + { 124 + 22 \xi \ov 3} [1,1] \
 .  \la{kan} \eea
Here  the three terms 
are the  contributions of the  FFB, FFBB and FFFF graphs, respectively, and
\be
[a,b]\equiv I(a) I(b) =
\mu^{2\ep}
\int { d^d p d^d q  \ov (2\pi)^d} {1 \ov ( p^2 + a) (q^2 + b)   }
 \la{deq} \ ,
   \ee
where $I(a)$ was defined in  \rf{oop}. Thus $[a,b]$ contains the 
${ 1\ov  \ep^2}+ { 1\ov  \ep} $ divergences.
%
%
%
The four terms in the last  FFFF paranthesis represent the
contributions of the $(\bar \theta \cD \theta)^2$ term in \rf{yy1},
of the term with $\G^{ab}$  in \rf{yy1},
of the term with $\G^{a'b'}$  in \rf{yy1}
and of the last term in \rf{yyo1}, respectively.
We find  that the   $[1,2]$ terms in \rf{kan} cancel,
but  there is no  cancellation of
the remaining terms, contradicting the  expected conformal invariance
of the theory.

In general, one may expect that in a (globally) supersymmetric theory the
 regularization of the fermionic and bosonic
parts of the action
should be done in some  consistent way. For bosons we used
dimensional regularization,
 and the
cancellation of $1 \ov \ep^2$ pole in \rf{dii},\rf{hjl} ensured also that the remaining
$1 \ov \ep$ pole had rational coefficient. Even if we  would  manage to cancel the
$1 \ov \ep^2$ pole in the fermionic contribution we would then need
some  sort of dimensional regularization producing $d$-dependent
 coefficients so that $1 \ov \ep$ pole had rational coefficient
   to be able to cancel its bosonic counterpart. Which kind of regularization is
   to be used to ensure that is unclear at the moment. The required
  rationality of the coefficient of the $1 \ov \ep$ pole   suggests that
   the   coefficients of $[1,4]$ and $[1,1]$  terms in \rf{kan}
   should, like coefficient of the
  $[1,2]$ term, be  separately  equal to zero.

Extracting the non-trivial finite  part  with 3 propagators
  contained in  the FFB  contribution we find that it is given by the integral
  \rf{jpy} (the integral \rf{df} does not   appear) but its coefficient is
  also  gauge ($\xi$) dependent
  \be     X_{\rm fin}= (4 + 2  \xi)  I(\sqrt 2, 1)   \la{fe}  \ee
 This gauge dependence of the UV  divergences
 and of the finite part  which should not be  present in the on-shell effective action
 is indicating a problem with
  maintaining $\k$-symmetry at the quantum level in the
  computational   prescription we have used.


\bigskip

Given the unsatisfactory result we found
 in the $\theta^1 = k \theta^2$ gauge we decided  to redo  the computation
 in a light-cone $\k$-symmetry gauge    which is the direct  analog
 of the usual $\G_+ \theta^I=0$ gauge
 in which the flat-space GS action becomes quadratic.
 The quadratic and quartic fermionic terms in the \adss action
 in this gauge are listed   in Appendix D.  Using a similar
 computational prescription as described above  we have obtained
 the  following  counterparts of eqs.
 \rf{kan}  and  \rf{fe}
\bea
&& X_\infty =
\Big(8[1,2]  + 12  [1,4]+  8 [1,1]\Big) +\Big(-8[1,2]  - 12 [1,4]\Big) \cr
&& \ \ \ \  -   { 4 \ov 3} (36 +   11  + 24
%
%
)   [1,1] =    - \frac{260}{3} [1,1]
%
%
\ ,  \la{akn} \\
&& \ \      X_{\rm fin}= 4 I(\sqrt 2, 1)  \ .   \la{fke}  \ee
Here  the three structures   in
$X_\infty$ are again  the contributions of the   FFB, FFBB and FFFF terms.
%
%
The three terms in the last paranthesis represent the contributions
of the \rf{yyr} term, of the first term in \rf{oyg} in the
${\cal M}^2$ term in \rf{qqu} and of the second and third terms in \rf{oyg}
in \rf{qqu}, respectively.

Again, the divergences do not appear to cancel\foot{Power-like
divergences have been eliminated in both equations \rf{akn} and
\rf{kan} due to our regularization scheme. It is, however, interesting
to note that in a cutoff-based regularization scheme the power-like
divergences appearing in the light-cone gauge are milder than those in
the $\theta^1=k\theta^2$ gauge. In particular, quartic divergences
appear to be absent in the former gauge.}
but one piece of good news  is  that
 the total coefficients not only of  the  $[1,2]$  but also of the  $[1,4]$ structures
 vanish  just as they did   in the $k=1$  ($\xi=0$) gauge  in \rf{kan}.
 Moreover,  the finite term in \rf{fke}
 is  exactly the same as \rf{fe}  in the  $k=1$ gauge.

\bigskip 

   Assuming  that our computational procedure
   can be corrected  so that the results   in  the two gauges fully agree
   with all divergences cancelling out   and the finite part still given by
     \rf{fke},\rf{jpy}  then  that would result in the
       fermionic  contribution 
     to 
     $a_2$ in \rf{ns}   with  $c_F=-2 $, i.e.\foot{In translating the result of computation of the 
     fermionic loop contribution  into the value of $a_{2F}$   we again 
     take into account the overall  sign change in the 2-loop term as required by 
     \rf{idi}.}
     \be \la{pre}
a_2=a_{2B} + a_{2F}  = {1 \ov \pi}  (1 - 2)   \KK   = -{1 \ov \pi}     \KK  
\approx  -0.29156\ . \ee
 Remarkably, this is in  good agreement with the numerical value \rf{kkkk}
found in \ci{kleb}  and reproduces exactly the value  of $a_2$ 
found recently from   the analytic solution of the BES equation in \ci{bkk}.

\section{Concluding remarks }

In this paper we initiated the study of 2-loop  quantum corrections
in \adss   string theory  on a particular example
of  the expansion near a simple ``homogeneous''  classical string solution.
We used conformal  gauge  for the 2d diffeomorphisms
and considered  two different choices (``covariant'' and ``light-cone'')
for the $\k$-symmetry gauge.

While we did not   manage  to completely  sort out the expected cancellation of
2-loop  UV divergences between the bosonic and the fermionic contributions,
our computation  revealed  the special transcendental structure of the finite  term
in the 2-loop effective action that determines the
next-to-next-to-leading order  coefficient  $a_2$ in the strong-coupling
expansion of the cusp
%
%
%
anomalous dimension on the gauge theory side of the AdS/CFT correspondence.
We expect that an improved  version of our
 computation\foot{One may try to redo the same  computation  using
a different fermionic parametrization of the \adss  action (e.g., like the one
employed  in \ci{frol}).
It would  be interesting also
 to attempt to do  a similar computation  by starting with
the Berkovits formulation \ci{ber} of the \adss action.}
that will resolve the technical problems
of apparent gauge dependence and non-cancellation of part of the
divergences
will not  change our conclusion about the  finite part
determining the value \rf{pre}  of the  coefficient   $a_2$ in \rf{uuu}.

The reason why we have more confidence in our result for the finite
rather than divergent part
of the 2-loop contribution is that, as explained in section 3,  the former
is determined only  by the quadratic fermionic terms in the \adss
superstring action
\rf{qaq}, while the latter depends essentially also on the complicated
quartic fermionic terms \rf{qqu}.\foot{There is of course an issue of
apparent gauge dependence of the finite part \rf{fe} in the
$\theta^1=k \theta^2$ gauge, but given that we got the same finite
results in the two very different gauges -- $\theta^1= \theta^2$ and
the light-cone gauge -- we are inclined to speculate that there is
some problem with the computation in the $k\not=1$ gauge.}


The  result \rf{ns},\rf{pre}  for  the 2-loop coefficient $a_2$ 
 suggests the following observation.
It is interesting to note that the first three terms in the strong
coupling expansion of the cusp anomalous dimension \rf{uuu}  hint at
a systematic expansion in polygamma functions. Indeed, $a_1$ in \rf{uuu} can be written as 
$a_1=-\frac{3}{2\pi}(\Psi(1)-\Psi({1 \ov 2}))$ and  $a_2$  
is proportional to the Catalan's constant 
$\KK$ \rf{cata}   which 
 contains  only the values of the first derivative
of the digamma function $\Psi(z)$.\foot{One may wonder
    if the actual mechanism of
    cancellation of UV divergences may leave behind a
  finite  piece
 containing
  $\ln 2$ terms. The presence of such $\ln 2$ terms could
   be in conflict
   with the ``transcendentality
 principle'' assuming one  extends it from weak-coupling  \ci{lip,bes}
 to a strong-coupling expansion. We thank M. Staudacher  for this remark.}
It is therefore tempting to conjecture that the coefficient $a_{n+1}$
appearing at 
order $\lambda^{-n/2}$ in the strong coupling expansion in \rf{sca},\rf{uuu}
 will be a
combination of values of derivatives  $\Psi^{(n)}(z)$ at rational arguments. 
A potentially related structure may follow from the strong coupling
expansion of the BFKL kernel which at weak coupling expresses the
finite spin twist-2 anomalous dimensions as an expansion in
derivatives of the digamma function (see \cite{KLRSV} for a comparison
between this approach and the Bethe ansatz predictions).

Similar 2-loop computations can also be done for some other special
string solutions, for example, for the 2-spin $(J_1,J_2)$ solution in
$S^5$. This solution further simplifies
 in the limit $J_1 \gg J_2$,  and  the 1-loop
correction vanishes \ci{mtt3};  the same is expected
\ci{mh}   to happen at the two  (and higher) loop level. The methods
of the present paper allow one to verify this.

\bigskip

\section*{Acknowledgments }

We are  grateful to   G.~Arutyunov, A. Belitsky, 
 L.~Dixon, S.~Frolov, I.~Klebanov, G. Korchemsky, 
T.~McLoughlin, R.~Metsaev, A.~Stasto, M.~Staudacher, X.~Yin and D.~Zanon
for many useful communications and discussions. R.R.  also acknowledges
the support of the National Science Foundation under grant
PHY-0608114. A.T. thanks the Physics Department of  The Ohio State
University for its support. A.A.T. acknowledges  the support of
the PPARC, INTAS 03-51-6346, EC MRTN-CT-2004-005104   and the RS
Wolfson award. Part of this work was done while A.A.T. was a
participant of the ``String and M Theory approaches to particle
physics and cosmology'' workshop at the Galileo Galilei Institute
for Theoretical Physics in Florence.
\bigskip

\renewcommand{\theequation}{A.\arabic{equation}}
\renewcommand{\thesection}{A}
 \setcounter{equation}{0}
\setcounter{section}{1} \setcounter{subsection}{0}

 \section*{Appendix A:   \adss  superstring  Lagrangian       }

The starting  point of the 2-loop computations in this paper is the
type IIB Green-Schwarz
\adss superstring action  $I=  \int d^2 \s \ {\cal L}
$   which is the sum of the ``kinetic'' and ``Wess-Zumino''
 term   \ci{mt}
\begin{eqnarray}
{\cal L}&=&{\cal L}_{Kin} + {\cal L}_{WZ}=
{ \sql \ov 2 \pi} \bigg[-\frac{1}{2}\sqrt{-h}h^{\alpha\beta}L_\alpha^A
L_\beta^A-2i\epsilon^{\alpha\beta} \int_0^1ds\, L_{\alpha s}^A \ {\rm s}^{IJ}
{\bar\theta}^I\Gamma^A L_{\beta s}^J
\bigg]
 \ .
\label{Lsusy}
\end{eqnarray}
The explicit form of this action to quartic order in $\theta$
(which is sufficient for our present purpose)  was presented  in \ci{mt}.
The exact solution of the Maurer-Cartan equations for the supervielbeine
was given in \ci{kal} (see also \ci{mett}).
The  \adss supersymmetry algebra
and thus the resulting  string action
of \ci{mt}
 can be rewritten in
terms of 10d Dirac matrices making it independent of a choice of a particular
representation of $\G^A$  \ci{hats} (see  also \ci{dgt,ft1}  and \ci{cal}).\foot{This
``10d covariant'' form of the action naturally comes out of the general form of
GS action in type IIB supergravity background \ci{howe}
once one specifies  the curvature
 and the 5-form field to their \adss values.}

 In the above expression $I,J=1,2$,\ ${\rm s}^{IJ}= (1,-1)$,
  $L_\a^A = (L_{\a s}^A)_{s=1}$       and
\begin{eqnarray}
L_{\alpha s}^A&=&\partial_\alpha x^\rho
e_\rho^A(x) -4i{\bar\theta}^I\Gamma^A \big[
\frac{\sinh^2({s\ov 2}{\cal M})}
{{\cal M}^2}\big]_{IJ}  \ D_\alpha\theta^J\ , \ \ \ \ \ \
L_{\beta s}^J=\big[\frac{\sinh (s{\cal M})}
{{\cal M}}D_\beta\theta\big]^J, \la{jk}
\end{eqnarray}
\begin{eqnarray}
D\theta^I={\cal D}\theta^I-\frac{i}{2}\epsilon^{IJ}
e^A\Gamma_*\Gamma_A\theta^J
~ \ ,  \ \ \ \    ~~~
{\cal D}\theta^I=d\theta^I+\frac{1}{4}\omega^{AB}\Gamma_{AB}
\theta^I\ , \ \ \ \ \ e^A=dx^\mu~e_\mu^A(x)
\label{des}
\end{eqnarray}
\begin{eqnarray}
({\cal M}^2)^{IL}=-\epsilon^{IJ}
\Gamma_* \Gamma^A\theta^J{\bar\theta}^L\Gamma_A
+\frac{1}{2}\epsilon^{LK}(\Gamma^{ab}\theta^I{\bar\theta}^K\Gamma_{ab}
\Gamma_* -
\Gamma^{a'b'}\theta^I{\bar\theta}^K\Gamma_{a'b'}\Gamma_*') \ . \la{oyg}
\end{eqnarray}
Here  $D^{IJ} D^{JK} \theta^K  =0$.
The indices run as follows
$$
\m,\n=0,1,2,...,9; \ \ \ \ A=(a;a')~; \ \ \ \ \ \ \ \
a,b=0,1,2,3,4~~; ~~~a',b'=5,6,7,8,9~~
$$
For Dirac matrices we    used  the notation from  \ci{cal}
\begin{eqnarray}
&&
\Gamma_*=i\Gamma_0\Gamma_1\Gamma_2\Gamma_3\Gamma_4
~, ~~~
\Gamma_*'=i\Gamma_5\Gamma_6\Gamma_7\Gamma_8\Gamma_9
~, ~~~
\Gamma_*\Gamma_*'=-\Gamma_*'\Gamma_*= \Gamma_{11}\ , \la{ga} \\
 &&
\Gamma_*^2=- \Gamma_*'^2=1\ , \ \ \ \ \
\Gamma_{11}= -\G_{0123456789} \ , \ \ \ \
\Gamma_{11}^2 = 1  \  . \la{gga}
\end{eqnarray}
Here $\G_A$ are $32 \times 32$ Dirac matrices,
$\G_{(A} \G_{B)}=\eta_{AB}=(-1,+1,...,+1)$,
and
 $\Gamma_{11}$
 defines the 10d  chiral projectors.
 We also assume the standard hermitian conjugation rule
 for fermions: $(\psi \chi)^\dagger = \chi^\dagger \psi^\dagger$.

 In the  type IIB  string action the
fermions are Majorana-Weyl of the
same chirality, e.g.,  $\theta^I = \G_{11} \theta^I$.
The Majorana  condition
\begin{eqnarray}
\bar \theta = \theta^T \cC \ , \ \ \ \ \ \
 \bar \theta \equiv  \theta^\dagger \G^0 \ , \ \ \ \ \
{\cC}^T=-{\cC}~~, ~~~~~~~
\Gamma_A=-{\cal C}^{-1}\Gamma_A^T{\cal C}~. \la{pr}
\end{eqnarray}
can be  solved by  choosing  $\cC= \G^0$ and thus  having
 $\theta$  real.\foot{For 10d Majorana fermions of the same chirality
  $\bar \psi_1   \G_{A_1...A_n}  \psi_2$
 is  non-zero  for $n$=odd   and is symmetric in $\psi_1,\psi_2$
  for $n=3,7$ and antisymmetric  if $n=1,5,9$.}
  In the  specific representation of $\G$-matrices used in \ci{mt,cal}
$\G_{11}=I_{16} \times \sigma_3$, so that
``left''  spinors satisfying  $\theta^I= \G_{11} \theta^I $ have
lower 16 components equal to zero.
The final result of our computation should not depend on a
choice of a particular representation of $\G_A$ and $\cC$.



To quartic order in fermions
 the fermionic part of  (\ref{Lsusy})  is
   ($\LL= \LL_B + \LL_F, \  \  \LL_F= \LL_{F2} + \LL_{F4} + ...
     $)
\begin{eqnarray}
{ 2 \pi \ov \sql }\LL_{F2}&=& {i }
(\eta^{\alpha\beta}\delta^{IJ}
-\epsilon^{\alpha\beta}{\rm s}^{IJ}) {\bar\theta}^Ie\llap/{}_\alpha
D_\beta\theta^J\cr &=& {i} (\eta^{\alpha\beta}\delta^{IJ}
-\epsilon^{\alpha\beta}{\rm s}^{IJ})
{\bar\theta}^Ie\llap/{}_\alpha \big[
\delta^{JK} {\cal D}_\beta
-\frac{i}{2}\epsilon^{JK}\Gamma_*e\llap/{}_\beta\big]\theta^K \ , \la{qaq}
\end{eqnarray}
\be
{ 2 \pi \ov \sql }\LL_{F4}= (\eta^{\alpha\beta}\delta^{IJ}-\epsilon^{\alpha\beta}{\rm s}^{IJ})
\big[
\frac{i}{12}{\bar\theta}^Ie\llap/{}_\alpha
{\cal M}^2_{JK}D_\beta \theta^K  + \ha
({\bar\theta}^K\Gamma^A D_\alpha\theta^K)
({\bar\theta}^I\Gamma_A D_\b \theta^J) \big]  \la{qqu}
\ee
Here we used the conformal gauge $ \sqrt{-h}h^{\alpha\beta} = \eta^{\a\b}$ and
 \be
e\llap/{}_\a =  e^A_\a \G_A \  , \ \ \ \ \ \ e^A_\a = e^A_\r \del_\a x^\r \ ,
\ \ \ \
\ \ \ \ \
 \   {\cal D}_\b=
  \partial_\beta+\frac{1}{4}\omega_\beta{}^{AB}\Gamma_{AB} \ ,\ \ \ \ \ \
   \omega_\a{}^{AB}= \omega_\rho {}^{AB}\del_\a x^\r \ .
  \la{der}\ee
    The metric, vielbeine and spin connection are those following
from the \adss metric (\ref{metric}).
In particular, the non-zero  background values  are (see
\rf{metric},\rf{jio},\rf{pip})\foot{We recall that the $AdS_5$ and $S^5$ coordinates
in \rf{ads},\rf{ssm}  are
labeled as $0,1,2,3,4$ and $5,6,7,8,9$.}
\bea \la{ljl}
e\llap/{}_0 &=& { \k \ov \sqrt 2}  (\G_8 + \G_9) \ ,\ \ \ \ \
e\llap/{}_1 = - { i \k \ov \sqrt 2}  (\G_8 - \G_9) \ ,\ \ \ \   e^A_\a e_{A \b} = -
\k^2  \eta_{\a\b} \ ,          \\
\omega_0{}^{AB}\Gamma_{AB}&=& \sqrt 2 \k  \G_7 ( \G_8 - \G_9)=  2i
   \G_7 e\llap/{}_1 \ , \ \ \ \ \ \ \
\omega_1{}^{AB}\Gamma_{AB}=- i\sqrt 2 \k  \G_7 ( \G_8 + \G_9)
= -2i     \G_7  e\llap/{}_0 \   . \no
 \eea
Let us  list the general expressions for the projected
vielbeine $e^A_\a = e^A_\r \del_\a x^\r$
 for the \adss metric in \rf{ads},\rf{ssm}
\be\la{ev}
e_\alpha^0=\frac{1+\frac{1}{4}z^2}{1-\frac{1}{4}z^2}\partial_\alpha t
~, \ \ \ \ \ ~~~~~~~
e_\alpha^k=\frac{1}{1-\frac{1}{4}z^2}\partial_\alpha z^k~,\ \ \ \ \ \ \ \ \   ~~~k=1,2,3,4
\ee
\be
e_\alpha^5=\frac{\sqrt{1-y^2}}{\sqrt{1-x^2-y^2}}
\partial_\alpha x+\frac{xy}{\sqrt{(1-y^2)(1-x^2-y^2)}}
\partial_\alpha y
~, \ \ \ \ ~~~~~~~
e_\alpha^6=\frac{\partial_\alpha y}{\sqrt{1-y^2}}
\ee
\be
e_\alpha^7=\sqrt{1-x^2-y^2}\partial_\alpha\psi
\ ,\ee \be
e_\alpha^8=\sqrt{1-x^2-y^2}\cos\psi\partial_\alpha\phi_2
\ , \ \ \ \ \ \
e_\alpha^9=\sqrt{1-x^2-y^2}\sin\psi\partial_\alpha\phi_3  \ .
\ee
The  Lorentz  connection  satisfying  $ \ep^{\a\b} ( \partial_\a e_\b^A
+\omega_\a{}^A{}_B e_\b^B)
=0$
($\omega_\alpha^{AB}\equiv \omega_\r{}^{AB} \partial_\alpha x^\r
= - \omega_\alpha^{AB} $) is
\be\la{we}
\omega_{\alpha}{}^{0i}
=\partial_\alpha t\,\frac{z^k}{1-\frac{1}{4}z^2}
~, ~~~~~~~
\omega_{\alpha}{}^{kn}
=-\frac{1}{2}\frac{z^k\,\partial_\alpha z^n
 -  z^n\,\partial_\alpha z^k }{1-\frac{1}{4}z^2}
\ , \ \ \ \ \ \ \ \
\omega_\alpha{}^{56}
=
-\frac{y}{\sqrt{1-y^2}} e_\alpha^5
\ee
\be
\omega_\alpha{}^{57}
=
\frac{x\partial_\alpha\psi}{\sqrt{1-y^2}}
\ , \ \ \ \ \ \ \ \ \ \ \ \
\omega_\alpha{}^{58}
=
\frac{x\cos\psi\,\partial_\alpha\phi_2}{\sqrt{1-y^2}}
\ee
\be
\omega_\alpha{}^{59}
=
\frac{x\sin\psi\,\partial_\alpha\phi_3}{\sqrt{1-y^2}}
\ , \ \ \ \ \ \ \ \
\omega_\alpha{}^{67}
=
\frac{y\sqrt{1-x^2-y^2}\partial_\alpha\psi}{\sqrt{1-y^2}}
\ee
\be
\omega_\alpha{}^{68}
=
\frac{y\sqrt{1-x^2-y^2}\cos\psi\,\partial_\alpha\phi_2}{\sqrt{1-y^2}}
\ , \ \ \ \ \ \ \ \ \
\omega_\alpha{}^{69}
=
\frac{y\sqrt{1-x^2-y^2}\sin\psi\,\partial_\alpha\phi_3}{\sqrt{1-y^2}}
\ee
\be
\omega_\alpha{}^{78}=
{\sin\psi\,\partial_\alpha\phi_2}
\ , \ \ \ \ \ \ \
\omega_\alpha{}^{79}=
-{\cos\psi\,\partial_\alpha\phi_3} \ .  \la{pip}
\ee

\renewcommand{\theequation}{B.\arabic{equation}}
\renewcommand{\thesection}{B}
 \setcounter{equation}{0}
\setcounter{section}{1} \setcounter{subsection}{0}

 \section*{Appendix B:  $\k$-symmetry  gauge fixing: $ \theta^1 = k \theta^2$  gauge  }

One natural gauge choice (used also in
\ci{dgt,ft1,ft3})
in the present case is \foot{This gauge is singular if one expands near a
null geodesic   but is regular if the string background has
 both $\tau$ and $\s$ dependence.}
\be \theta^1=\theta^2 \equiv \theta \ . \la{y} \ee
Then for  the relevant (to  2-loop order)  quartic  terms  in the fermions
one finds
\begin{eqnarray}
L_{\alpha s}^A&=&\partial_\alpha x^\rho
e_\rho^A-2is^2{\bar\theta}\Gamma^A{\cal D}_\alpha\theta
+
\frac{s^4}{12}{\bar\theta}\Gamma^A
(-\Gamma^{ab}\theta{\bar\theta}\Gamma_{ab}
\Gamma_*+
\Gamma^{a'b'}\theta{\bar\theta}\Gamma_{a'b'}\Gamma_*')
\Gamma_*\Gamma_B\theta\ e^B_\alpha\cr
&=&\partial_\alpha x^\rho e_\rho^A
-2is^2{\bar\theta}\Gamma^A{\cal D}_\alpha\theta
+
\frac{s^4}{12}{\bar\theta}\Gamma^A
(-\Gamma^{ab}\theta {\bar\theta}\Gamma_{ab}+
\Gamma^{a'b'}\theta {\bar\theta}\Gamma_{a'b'}
)
\Gamma_B\theta \ e^B_\alpha\cr
{\rm s}^{IJ}{\bar\theta}^I\Gamma^A
L_{\beta s}^J&=&-is{\bar\theta}\Gamma^A\Gamma_*\Gamma_B\theta\ e^B_\beta
-\frac{2s^3}{3}{\bar\theta}\Gamma^A\Gamma_*\Gamma^B\theta
{\bar\theta}\Gamma_B{\cal D}_\beta\theta \ .
\end{eqnarray}
As a result,
 the ``kinetic'' and ``WZW''  parts of \rf{Lsusy} become
(to order $\theta^4$)
\begin{eqnarray}
{ 2 \pi \ov \sql }
\LL_{\rm Kin}
&=&\ \eta^{\alpha\beta} \big[   -\frac{1}{2}
\partial_\alpha x^\m\partial_\beta
x^\n G_{\m\n}(x)
+2i e_\alpha^A{\bar\theta}\Gamma_A{\cal
D}_\beta\theta
+2
{\bar\theta}\Gamma^A{\cal D}_\alpha\theta{\bar\theta}\Gamma_A{\cal
D}_\beta\theta \cr
&&+\frac{1}{12}e_\alpha^Ae_\beta^B \
{\bar\theta}\Gamma_A
(\Gamma^{ab}\theta {\bar\theta} \Gamma_{ab}-
\Gamma^{a'b'}\theta {\bar\theta} \Gamma_{a'b'}
)
\Gamma_B \theta \big]  \ , \la{yy}
\end{eqnarray}
\begin{eqnarray}
{ 2 \pi \ov \sql } \LL_{\rm WZ}&=&
\epsilon^{\alpha\beta} \big[
- e^A_\a  e^B_\b {\bar\theta}\Gamma_A\Gamma_*\Gamma_B\theta
 + \frac{i}{3}e^A_\a {\bar\theta}\Gamma_A\Gamma_*\Gamma_B\theta
{\bar\theta}\Gamma^B{\cal D}_\b \theta
-ie^A_\a {\bar\theta}\Gamma_B\Gamma_*\Gamma_A\theta
{\bar\theta}\Gamma^B{\cal D}_\b \theta \big]
\nonumber
\\
&=&\epsilon^{\alpha\beta} \big[ -e^A_\a e^B_\b
 {\bar\theta}\Gamma_A\Gamma_*\Gamma_B\theta
+\frac{4i}{3}e^A_\a {\bar\theta}\Gamma_A\Gamma_*\Gamma_B\theta
{\bar\theta}\Gamma^B{\cal D}_\b\theta \big]   . \la{yyo}
\end{eqnarray}
We  used that
for the  ``left''  fermions
$\G_{11} \theta= \G_* \G_*' \theta = \theta$ and  also that
$\bar{\theta}\Gamma_B\Gamma_{*}\Gamma_A\theta=
-\bar{\theta}\Gamma_A\Gamma_{*}\Gamma_B\theta$.
The resulting action is the same as the quartic fermionic
action  found in eqs. (4.12)-(4.14) in \ci{mt}  upon restricting it to the gauge
\rf{y}.

One may also  consider a  more general gauge
(here $k$ is a real number)
\begin{equation}\la{ot}
\theta^1=\ k\ \theta^2 \ , \ \ \ \ \ \ \ \ \  \theta^2\equiv \theta \ .
\end{equation}
Then to $\theta^4$ order
\begin{eqnarray}
L_{\alpha s}^A&=&\partial_\alpha x^\rho e_\rho^A
-i(1+k^2)s^2{\bar\theta}\Gamma^A{\cal D}_\alpha\theta\cr
&+&
(1+k^2)^2\frac{s^4}{48}{\bar\theta}\Gamma^A (-\Gamma^{ab}\theta
{\bar\theta}\Gamma_{ab}+ \Gamma^{a'b'}\theta
{\bar\theta}\Gamma_{a'b'}
) \Gamma_C\theta e^C_\alpha \ , \cr
{\rm s}^{IJ}{\bar\theta}^I\Gamma^A L_{\beta
s}^J&=&(k^2-1)s \bar{\theta}\Gamma^A{\cal D}_\beta \theta
 -i k s {\bar\theta}\Gamma^A\Gamma_*\Gamma_B\theta e^B_\beta
 -k(1+k^2)
\frac{s^3}{3}{\bar\theta}\Gamma^A\Gamma_*\Gamma^B\theta
{\bar\theta}\Gamma_B{\cal D}_\beta\theta
\nonumber\\
&+& (k^4-1)\frac{i s^3}{24}\bar{\theta}\Gamma_A (-\Gamma^{ab}\theta
{\bar\theta}\Gamma_{ab}+ \Gamma^{a'b'}\theta
{\bar\theta}\Gamma_{a'b'}
) \Gamma_C\theta e^C_\beta\ .
\end{eqnarray}
As a result, \rf{yy} and \rf{yyo} are generalized to
\begin{eqnarray}
{ 2 \pi \ov \sql } \LL_{\rm Kin}
&=& \eta^{\alpha\beta} \big[   -\frac{1}{2} \partial_\alpha
x^\m\partial_\beta x^\n G_{\m\n}(x) +i (1+k^2)
e_\alpha^A{\bar\theta}\Gamma^A{\cal D}_\beta\theta
+\frac{(1+k^2)^2}{2}{\bar\theta}\Gamma^A{\cal
D}_\alpha\theta{\bar\theta}\Gamma^A{\cal D}_\beta\theta \cr
&-&\frac{(1+k^2)^2}{48}  e_\alpha^Ae_\beta^B
{\bar\theta}\Gamma_A (-\Gamma^{ab}\theta {\bar\theta} \Gamma_{ab}+
\Gamma^{a'b'}\theta {\bar\theta} \Gamma_{a'b'}
)
\Gamma_B\theta \big] \ ,  \la{yy1}
\end{eqnarray}
\begin{eqnarray}
{ 2 \pi \ov \sql } \LL_{\rm WZ}&=&  \ep^{\a\b} \big[
-i(k^2-1) e^A_\alpha \bar{\theta}\Gamma_A{\cal
D}_\beta\theta-ke^A_\a e^B_\b  {\bar\theta}\Gamma_A\Gamma_*\Gamma_B\theta\cr
&&~~~ +
 \frac{1}{6} i k(k^2+1)e^A_\a {\bar\theta}\Gamma_A\Gamma_*\Gamma_B\theta
{\bar\theta}\Gamma^B{\cal D}_\b \theta
- \ha i {k(k^2+1)} e^A_\a
{\bar\theta}\Gamma_B\Gamma_*\Gamma_A\theta {\bar\theta}\Gamma^B{\cal
D}_\b \theta
\big]
\nonumber\\
&=&\epsilon^{\alpha\beta} \big[
-i(k^2-1)e^A_\alpha \bar{\theta}\Gamma_A{\cal D}_\beta\theta-k e^A_\a
e^B_\b {\bar\theta}\Gamma_A\Gamma_*\Gamma_B\theta \cr
&& ~~~ +
\frac{2}{3}i k(k^2+1 )e^A_\a {\bar\theta}\Gamma_A\Gamma_*\Gamma_B\theta
{\bar\theta}\Gamma^B{\cal D}_\b \theta
\big] \ ,
\la{yyo1}
\end{eqnarray}
where we used that the term proportional to $k^4-1$ vanishes under
 antisymmetrization in $\a,\b$.
Note that if we rescale $\theta$ by  $ ({k^2+1 \ov 2})^{1/2}$
then \rf{yy1} will become equivalent to \rf{yy}
while \rf{yyo1}   will take the form
\begin{eqnarray}
{ 2 \pi \ov \sql } \LL_{\rm WZ}
&=&\epsilon^{\alpha\beta} \big[
-2 i{ k^2-1 \ov k^2 + 1} e^A_\alpha
\bar{\theta}\Gamma_A{\cal D}_\beta\theta-{ 2 k\ov k^2 + 1}
 e^A_\a
e^B_\b {\bar\theta}\Gamma_A\Gamma_*\Gamma_B\theta \cr
& +&
\frac{8i}{3} { k \ov k^2 +1 } e^A_\a {\bar\theta}\Gamma_A\Gamma_*\Gamma_B\theta
{\bar\theta}\Gamma^B{\cal D}_\b \theta \big]\ ,
\la{yyo2}
\end{eqnarray}
which reduces to  \rf{yyo} for $k=1$.

%
%
%

The fermionic propagator
 in the $\theta^1=k\theta^2$ gauge
corresponding to \rf{yy},\rf{yyo2}
(after the above rescaling of $\theta$ and after the rescaling
of momenta by $\k$, i.e. with  the same normalization as in   \rf{prop}) is
\bea
K^{-1} (q)&=&\frac{k^{-1}  + k  }{8\sqrt{2}\,(q^2+1)}
\bigg(~~~~\Big[\frac{k^{-1}}{2}(1-i)(q_0-q_1)+\frac{k}{2}(1+i)(q_0+q_1)\Big]
\Gamma_8\cr
&&~~
     +\Big[\frac{k^{-1}}{2}(1+i)(q_0-q_1)+\frac{k}{2}(1-i)(q_0+q_1)\Big]
\Gamma_9
     -i\sqrt{2} \,\Gamma_*\Gamma_8\Gamma_9\bigg){\cal C}
     \la{few}
\eea
where  $q$ is the 2d momentum and  ${\cal C}$ is the charge conjugation matrix.
Note that the  contribution of the
connection terms in ${\cal D}_\a$ to the propagator vanishes (cf. \rf{ljl}).

The propagator  is invariant  under $k \to k^{-1}$ combined with
the 2d parity transformation, i.e.
$q_1 \to - q_1$. The same transformation is also a symmetry of the interaction terms in
\rf{yy1},\rf{yyo1}.\foot{Note that the GS action \rf{Lsusy}
is not invariant under  $\theta^1 \to \theta^2$ due to:
 (i)
the presence of s$^{IJ}$ in the WZ term, and (ii) the presence of
$\epsilon^{IJ}$ terms in $D \theta$ and in ${\cal M}$ in \rf{oyg}.
The first reason is present already in flat-space  GS action and can be compensated by
2d parity transformation or $\ep^{\a\b} \to - \ep^{\a\b}$.
The second is due to the presence of a non-trivial RR background: each
$\epsilon^{IJ}$ factor is accompanied  by a factor of $\G_*$ (note that
 $\G'_*= \G_* \G_{11}$)
which is present due to coupling to self-dual $F_5$ field.
Thus reversing the sign of $F_5$  background corresponds to
 $\theta^1 \to \theta^2$ combined with 2d parity transformation.
 }

\renewcommand{\theequation}{C.\arabic{equation}}
 \setcounter{equation}{0}
\setcounter{section}{1} \setcounter{subsection}{0}

 \section*{Appendix C:  Cancellation of  2-loop corrections  in\\
  flat-space Green-Schwarz action in $\theta^1=\theta^2$ gauge}


To clarify the issue of cancellation of
power divergences in diagrams with fermion lines
it is useful to consider a similar 2-loop cancellation in  flat
space type IIB GS action \ci{gs}
(cf. \rf{Lsusy})
\be\la{fl}
I=  { 1 \ov 2 \pi \a'} \int d^2 \s \
 \bigg[-\frac{1}{2} (\del_\a x^\m  - i{\bar\theta}^I\Gamma^\m \del_\alpha\theta^I)^2
-i\epsilon^{\alpha\beta} {\rm s}^{IJ} \bt^I \G_\m \del_\b \theta^J\  ( \del_\a x^\m
- \ha i  {\bar\theta}^K \G^\m \del_\a \theta^K) \bigg]\  ,
 \ee
where we fixed the conformal gauge $\sqrt{-h}h^{\alpha\beta}= \eta^{\a\b}$.
Let us expand this action near the ``homogeneous'' classical
solution\foot{Since the above action depends on $x^\m$ only through its derivatives, the coefficients in
the expanded action will be constant.}
\be \la{ham}
x^\m = N^\m_\a \s^\a \ , \ \ \ \ \ \ \ \ \ \ \ \ \s^\a= (\tau, \s)\ ,   \ee
where  $N^\m_\a$ are constant vectors
(which we may formally allow to be  complex) assumed to satisfy
\be \la{nor}
\del_\a x^\m \del_\b x_\m  = \eta_{\m\n} N^\m_\a N^\n_\b = \cc \eta_{\a\b}
\ . \ee
Here $\cc$ is
a background-dependent  constant.
 The direct analog of our  $S^5$ background in \rf{jio} is the following choice
\be\la{hj}
 N^2_\a = {\k \ov \sqrt 2 } (1, -i)\ ,\  \ \ \ \ \ \
\ N^3_\a = {\k \ov \sqrt 2 } (1, i)\ , \ \ \ \ \ \  \ \ \  \cc= -\k^2\  ,  \ee
where $x^2,x^3$ directions are  analogs of $\phi_2$ and $\phi_3$ in  \rf{jio}.

 Let us   fix the $\k$-symmetry by the same condition as in \rf{y}:
 $\te^1=\te^2\equiv \te$.  Since ${\rm s}^{IJ}=(1,-1)$,  the contribution
 of the WZ term in \rf{fl} then vanishes.
  The resulting fermionic kinetic term will turn out to be  non-degenerate so this
  gauge is admissible.

 Setting $x^\m \to x^\m + \td x^\m$, we get the following
 action for the fluctuations $\td x^\m, \ \te$
 \be\la{fla}
\td I=  { 1 \ov 2 \pi \a'} \int d^2 \s \
 \bigg[-\frac{1}{2} (\del_\a \td x^\m  - 2i{\bar\theta}\Gamma^\m \del_\alpha\theta)^2
+ 2i {\bar\theta}\gamma^\a \del_\alpha\theta \bigg]
\  ,
 \ee
where
\be \la{jg} \gamma_\a \equiv  N^\m_\a \G_\m \ , \ \ \ \ \ \ \ \ \ \ \ \ \ \ \Gamma_{(\m} \Gamma_{\n)}=
\eta_{\m\n} \ , \ \ \ \
\gamma_{(\a} \gamma_{\b)} = \cc \eta_{\a\b} \ .
  \ee
To this  action  we should add the  contribution of the conformal gauge ghosts
and
the $\kappa$-symmetry ghosts. The former is decoupled from the background
but the latter is non-trivial.
 The  invariance of the GS action under  the $\kappa$-symmetry
 $\delta \te^I=
 (\del_\a x^\m  - i{\bar\theta}^J\Gamma^\m \del_\alpha\theta^J)\G_\m \kappa^{\a I}$
 (here the spinor parameter
 $\k^{\a 1}$ is selfdual and $\k^{\a 2}$ -- antiselfdual in 2d vector index  $\a$)
 leads in the $\te^1=\te^2$ gauge
 to an ultralocal ghost action\foot{The  conformal gauge ghosts and the $\k$-symmetry
 ghosts decouple.}
 \be
I_{gh}(b,c) = { 1 \ov 2 \pi \a'} \int d^2 \s \  b^I
 ( N^\m_\a + \del_\a \td x^\m  - 2i{\bar\theta}\Gamma^\m \del_\alpha\theta)\G_\m\
   c^{\a I} \ . \la{hih}
 \ee
On general grounds,
one  should expect that the total string  partition function  should be trivial
 despite the non-linearity  of the action \rf{fla}. Indeed,
 we could have fixed  first the
 conformal gauge $x^+= p^+ \tau, \ \G_+ \theta^I=0$  in which the GS  action
 \rf{fl} becomes quadratic  and then choose the background \rf{ham} in the $x^2,x^3$
 directions transverse to $(x^+,x^-), \ x^\pm = x^0 \pm x^1$. Since we are
 expanding near an  on-shell
 background, the partition function
 should be  gauge-independent, i.e.  still trivial.

 Let us note that the resulting theory \rf{fla}
    is formally non-renormalizable:
 the fermionic kinetic term   is linear in 2d momentum while
 fermionic  interactions contain derivatives. This is a reflection
 of the absence of the (non-unitary)
 $\del \bar \theta \del \theta $ kinetic term in the GS action
 (i.e. of the degeneracy  of the  corresponding superspace sigma model metric).
 Thus we should expect divergences with
 higher  powers of the UV cutoff (in an appropriate
 covariant regularization); the triviality of quantum
 corrections requires  cancellation
  of all divergences, and,
  in particular,  the absence of logarithmic divergences.

Let us first consider the 1-loop approximation. Counting
non-trivial $\ha \ln \det (-\del^2)$ contributions  one gets
10 from bosons, -2 from conformal ghosts and  $ - \ha \times 16=8$ from
one  MW fermion $\theta$; this checks that the total effective number of degrees
of freedom is 0. In addition, there is a quadratic divergence  proportional to
$\ln \cc$  coming from the $\theta$-determinant ($(\g^\a \del_\a)^2
= \cc \del^2$). It is cancelled  by the 1-loop  contribution
 of the $\k$-ghosts in \rf{hih} ($\int d^2 \s\ b^I \g_\a
c^{I\a} + ...$).\foot{Similar cancellation applies to the $p^+$-dependence
in lightcone gauge.}

To compute the 2-loop contribution it is useful first to
 transform the action \rf{fla},\rf{hih}
  into an equivalent  but simpler-looking 2-d dual
(or ``T-dual'')  form\foot{A similar transformation was used in
 \ci{kalt}.}
 by introducing  two  auxiliary fields $L^\m_\a$ and  $P_\m^\a$
  and writing the total fluctuation action as
  \bea
I_{tot}=
  { 1 \ov 2 \pi \a'}\int d^2 \s \bigg[ &-& \ha (L^\m_\a)^2  +
   2i {\bar\theta}\gamma^\a \del_\alpha\theta\cr
   &+&
  b^I (\g_\a  + L^\m_\a \G_\m ) c^{I\a}
+ P_\m^\a [ L^\m_\a -( \del_\a \td x^\m  - 2i{\bar\theta}\Gamma^\m \del_\alpha\theta) ]
\bigg]  \ . \la{to} \eea
Integrating first over $\td x^\m $ (implying
$P_\m^\a = \ep^{\a\b} \del_\b y_\m$  where $y^\m$ is a
``2-d dual'' of  $\td x^\m$)  and then over $L^\m_\a$
results in
\bea \la{toj}
\td I_{tot}=
  { 1 \ov 2 \pi \a'}\int d^2 \s \bigg[ &-&
  \ha (\del_\a y^\m  + \ep_{\a\b} b^I  \G^\m  c^{I\b}) ^2  + 2i {\bar\theta}\gamma^\a \del_\alpha\theta
  - 2i \ep^{\a\b} \del_\b  y^\m  {\bar\theta}\Gamma_\m \del_\alpha\theta
  \cr
 &+& b^I \g_\a c^{I\a}  +  \ep_{\a\b} \del^\b y^\m  b^I \G_\m  c^{I\a}\bigg]
  \ . \eea
This can be written also as
\bea
\td I_{tot}=
  { 1 \ov 2 \pi \a'}\int d^2 \s \bigg[ &-&
  \ha (\del_\a y^\m)^2  + 2i {\bar\theta}\gamma^\a \del_\alpha\theta
  +  b^I \g_\a c^{I\a} \cr
   &-& 2i \ep^{\a\b} \del_\b  y^\m  {\bar\theta}\Gamma_\m \del_\alpha\theta
  + \ep_{\a \b } \del^\a y^\m  b^I  \G_\m  c^{I\b}
  + \ha  (b^I  \G^\m  c^{I\a})^2 \bigg]\ . \la{tokj}
  \eea
An  advantage of this form of the action is the absence of the $\theta^4$
and $bc \theta^2$ terms at
 the price of the appearance of (simpler) $(bc)^2$ term.\foot{To make the structure of possible cancellations more transparent it might be
useful to replace the (anti)selfdual ghost
 $c^{I\a}$   with   two {\it commuting} ghost spinor fields (the associated Jacobian is
background-independent):
$
c^{1\a} = ( \eta^{\a\b} + \ep^{\a\b}) \del_\b \vt^1 \ , \ \
c^{2\a} = ( \eta^{\a\b} - \ep^{\a\b}) \del_\b \vt^2 \ .
$
That way it may be possible to show the cancellation  of corrections  between
loops of $\theta$
and  loops of $(b,\vt^I)$ to all orders.
We will not pursue this here.}

Then
 the only 2-loop diagram involving $\theta$
  is then of type (a) in Figure 1 where
one line is bosonic and two lines are fermionic.
Because of the properties of $\g_\a$ in
\rf{jg}  the propagator for the Majorana-Weyl 10d spinor $\te$
 is essentially the same as for a 2-d fermion, i.e. is (in momentum
 representation)
${ p^\a \g_\a \ov p^2}.$
Then  the non-trivial  contribution  (from the diagram on Figure 1(a))
 to the
2-loop effective action is proportional to
($V_2$ is the 2d volume factor)
 \be \la{ge}
 { V_2 \ov \cc^2}
  \int { d^2 p d^2 q \ov  ( 2 \pi)^4 }
{   \Tr ( \G^\m  p^\a \g_\a\G_\m  q^\b \g_\b) \ \ep^{\g\delta} p_\g (p+q)_\delta
\ep^{\g'\delta'} q_{\g'} (p+q)_{\delta'} \ov p^2 q^2 (p+q)^2  }
\ . \ee
Since\foot{The trace is taken with the  Weyl projector implied.}
 $\Tr ( \G^\m  p^\a \g_\a\G_\m  q^\b \g_\b)= - 10 \times 16\ \cc  (pq)$
we end up with (omitting the prefactor $ { V_2 \ov \cc}$)
\bea
&&
  \int { d^2 p d^2 q \ov  ( 2 \pi)^4 }
{   (pq)  [ (pq)^2 - p^2 q^2  ]
 \ov p^2 q^2 (p+q)^2  } =
 \int { d^2 p d^2 q \ov 4 ( 2 \pi)^4 }
 \bigg[ 1
 + 3 {p^2\ov q^2}  - 3  {(p+q)^2\ov q^2}  \cr
&& \ \ \ \ \ \  \ \  \ \
 +\  {q^2\ov (p+q)^2}
 +  {(p^2+q^2 +2 pq)^2\ov 2 p^2 q^2}  - \ha  {p^4\ov (p+q)^2 q^2} \bigg]
\la{gep}
\eea
where we factorized the integrand  and used the  symmetry under
$ p \to q$ as well as  Lorentz invariance of the integrand.
The above integral can be simplified further into
\be
 \int { d^2 p d^2 q \ov 4 ( 2 \pi)^4 }
 \bigg[  {p^2\ov q^2}
 + {q^2\ov (p+q)^2}  - \ha  {p^4\ov (p+q)^2 q^2} \bigg]  \ .
\ee
This integral is quartically divergent.
Applying  the dimensional regularization
(in combination  with an IR regularization  by a mass, see  \ci{leib})
we conclude that it does not contain any logarithmically divergent or finite parts, i.e. the
result  vanishes. The contribution of ghosts is also trivial in dimensional regularization.

Alternatively,   we may use an explicit regularization like an exponential cutoff
by inserting
 $e^{-{p^2\ov \Lambda^2}}$ for each momentum
integral.
Then we get for \rf{gep} (omitting the overall factor)
 \bea
  \int { d^2 p d^2 q d^2 k \ov  ( 2 \pi)^4 }\delta^{(2)} (p+q+k)
\frac{(k^2-p^2-q^2) \ [(k^2-p^2-q^2)^2-4 \ p^2 q^2]}{8 p^2 q^2
k^2} \  e^ {-\frac{1}{\Lambda^2} (p^2+q^2+k^2)}\eea
 Using the symmetry
of the integrand  under interchange of   $p,q,k$ we obtain
 \bea
 && \int { d^2 p d^2 q d^2 k \ov  4( 2 \pi)^4 }\delta^{(2)}(p+q+k)
\big[1-\frac{k^4}{2 p^2 q^2}+\frac{k^2}{p^2}\big]\ e^
{-\frac{1}{\Lambda^2}(p^2+q^2+k^2)}\cr  && =
  \int { d^2 p d^2 q  \ov  4( 2 \pi)^4 }
\bigg[1-\frac{(p+q)^4}{2 p^2 q^2}+\frac{(p+q)^2}{p^2}\big] \ e^
{-\frac{1}{\Lambda^2}(p^2+q^2+(p+q)^2)} \  .  \eea
Evaluating the integrals here we find that  the   first term in the bracket gives
$ \frac{\Lambda^4}{192 \ \pi^2}$
   while each of the  last two gives zero.

The result is thus simply a quartic divergence,
 which should then be  cancelled against the
local $\k$-symmetry ghost contribution so that the total
2-loop contribution to the effective action is trivial.
A careful check of this cancellation may require a
systematic  development
of the phase-space quantization of the GS action in the $\theta^1=\theta^2$
gauge (with all measure  factors taken into account).\foot{In general,
  local  measure may  not be fixed in the Lagrangian quantization;
that means  also power divergences can not be cancelled unless all
 local factors of ghosts and measure are included. For a  previous discussion of
 quantization of flat-space  GS action see, e.g.,
\ci{kaa}.}
The use  of dimensional regularization allows one to
by-pass this problem. This is the strategy we  adopt also
 in the curved-space case considered in this paper.





\renewcommand{\theequation}{D.\arabic{equation}}
 \setcounter{equation}{0}
\setcounter{section}{1} \setcounter{subsection}{0}

 \section*{Appendix D:  $\k$-symmetry  light-cone   gauge
 $\G_+ \theta^I=0$ }

The  flat-space GS action is known to simplify dramatically
in the  $\k$-symmetry  light-cone   gauge $\G_+ \theta^I=0$:  the quartic fermionic term
in it
vanishes.  It is  natural to expect that
 a  choice of  a similar  gauge   may also lead to important simplifications
  in curved space-time  case.
  In particular, at least part of
  power divergences may  then be absent. Below we
  shall  present the  details of the structure of the
  \adss   action in a   light-cone   gauge   $\G_+ \theta^I=0$
  needed for computing  the fermionic 2-loop   contribution
  discussed in section 3.

\subsection*{D.1  Vanishing of 2-loop correction in the expansion near null geodesic}

As a preparation for the 2-loop computation we are interested in
it is useful   first to  consider the expansion near
the simplest point-like string configuration: null geodesic  that goes around $S^5$.
Since this is a BPS configuration preserving 1/2  of supersymmetry
one expects to find that all world-sheet  loop contributions  to
the  sigma model partition function expanded near this
background   vanish, i.e. the ground-state energy should not receive quantum corrections.
 This is indeed easily verified   in the 1-loop approximation
where  choosing the  light-cone    $\k$-symmetry gauge
one gets 8 bosonic and 8 fermionic fluctuation modes  with equal mass \ci{mets,bmn,ft1}.
We have checked  explicitly that the same   is true also in the 2-loop
approximation where one no longer has a benefit of an effective 2d supersymmetry or
even manifest 2d
Lorentz symmetry
present in the ``1-loop''
 (i.e. ``plane-wave'') action.

We shall use conformal gauge   and consider the expansion of the superstring
action near the following sigma model solution  corresponding to the  metric
\rf{ads},\rf{ssm}:
$t= \k \tau, \ \ \p_2= \k \tau$   with all other angles being trivial.
It is actually useful to change the parametrization of the $S^5$ metric from
\rf{ssm} to  the one similar to \rf{ads}:
\be
(ds^2)_{S^5} =
\big(\frac{1-\frac{1}{4}y^2}{1+\frac{1}{4}y^2}\big)^2 d\p^2
+\frac{dy^n dy_n}{(1+\frac{1}{4}y^2)^2} \ , \ \ \ \ \  \ \ \ \ \ n=1,2,3,4 \ .
\la{adse}\ee
Then the  classical solution (which solves both the sigma model equations
and the conformal gauge constraints) is
\be\la{hip}
t= \k \tau,\ \ \ \ \ \ \
 \p= \k \tau,\ \ \ \ \  \ z_k=0, \ \ \ \ \ \ y_n=0 \ ,
 \ee
  and we   should thus
 expand the action
to quartic order in fluctuation fields $\td t=t- \k \tau, \ \td \p=\p- \k \tau, \ z_k,\ y_n$
and $\theta^I$   subject  to the  l.c. $\k$-symmetry gauge condition
$(\G_0 + \G_5) \theta^I=0$ (we label $\p$ as the 5-th coordinate).

Let us first make general  comments on the bosonic contribution.
The logarithmically divergent parts of the effective actions of the   decoupled $AdS_5$ and $S^5$
sigma models are each given  by the counterterm \rf{zme}   multiplying the $\del x \del x$
term.
For a symmetric space \rf{zme}  is proportional to the metric itself, so we get,
up to numerical coefficients, $(\a' R+ \a'^2 R^2 + \a'^3 R^3 + ...) G_{\m\n}(x) \del x^\m \del x^\n$.
Since the scalar curvatures of $AdS_5$ and $S^5$   here are opposite  in sign,
we conclude that the  divergence at one (or any odd) loop  is proportional to the {\it difference}
of the $AdS_5$ and $S^5$  classical  actions, while the divergence at
two (or any even) loop   is proportional to the {\it sum}
of the $AdS_5$ and $S^5$  classical  actions (i.e. to the total classical   string action).
The  difference of the $AdS_5$ and $S^5$  classical  actions
is non-vanishing on \rf{hip}, in agreement with the presence of 1-loop divergence
coming from 8 equal-mass bosonic modes; this divergence is of course cancelled by the fermions.
The   sum of the   $AdS_5$ and $S^5$  classical  actions
   vanishes on the solution \rf{hip}, so  we conclude  that the bosonic
part of the partition function  can get only  finite contribution
at two (or any  even number of) loops.

This is indeed what we have found by the  direct 2-loop computation:
 the bosonic  2-loop contribution happens to be completely trivial, i.e.
the 2-loop bosonic part of the effective action vanishes.\foot{Note that our
computation is
different from the discussions of near-BMN expansion in \ci{cal,frol}
where  a light-cone-type gauge was imposed
on the bosons. We instead  use the conformal gauge, with the conformal gauge
ghosts cancelling
the contribution of
2 massless   longitudinal modes ($\td t$ and $\td \p$) at 1-loop;
%
%
within our regularization scheme
the contribution of these modes also decouples at higher loops.}

As for the fermionic part, we found (using the l.c. gauge expansion)  that the
contribution of the diagram in Figure 1(a)   with two Yukawa FFB  vertices is
identically zero,
while the contributions of  the  FFBB and   FFFF  terms in Figure 1(b)
are proportional to the square of the simple massive
 tadpole integral\foot{We again set $\k=1$ by a rescaling of 2d coordinates/momenta.}
 $[1,1]$ in \rf{deq}  with the coefficients being, respectively,
 32 and -32.
 Thus  the total 2-loop term in the effective action    
expanded near the 
null geodesic is indeed zero.

Let us stress  that to arrive at this result we  used  dimensional regularization
 only in a
limited sense: all tensor algebra was done in $d=2$
  and  we  continued to $d < 2$  (to eliminate power divergences)  only  at the very end
  for the scalar integrals found after factorization of highest divergent parts of the integrands.
   If instead we have used  the standard dimensional regularization
   (i.e. have  assumed that $\langle p_\a p_\b\rangle ={1 \ov d}  \eta_{\a\b} \langle p^2\rangle$
 instead of  $\langle p_\a p_\b\rangle = \ha \eta_{\a\b} \langle p^2\rangle$)
  then
the contribution of the FFFF term would be $-  64  ( 1 - { 1 \ov d})$
and we would  be left with non-cancelled  ${1\ov \epsilon}$ divergences
(and a finite part).
This indicates that the standard dimensional regularization cannot be
applied to the GS action: it  breaks some of its symmetries
which results in non-trivial corrections to what should be a protected  BPS state.
This of course is not surprising given, in particular,
 the presence of the  WZ term in the GS
 action.

\subsection*{D.2  Expansion near  the $S^5$ solution in the  light-cone gauge }

The background \rf{ro}    selects two  spatial directions
 $x_8\equiv \p_2,\ x_9\equiv \p_3$
so  a natural   choice for the  l.c. gauge  condition
  that should produce
a non-degenerate fermionic propagator when one expands near \rf{ro}   is
$
[\G_0 +  {1 \ov \sqrt 2} (\Gamma_8+\Gamma_9) ] \theta^I =0$.
More generally, we may consider a ``rotated'' choice
$[\Gamma_0+\frac{1}{\sqrt{1+\z^2}}(\Gamma_8+\z \Gamma_9)]\theta^I  =0 $
where $\z $ is a gauge-fixing parameter.
The  result for the  effective action
does not depend on the value of $\z$:
 since   $\G_A$  have tangent-space indices
 this  = follows from rotational invariance of the action  in the tangent space.
 In what follows  we  shall  choose the simplest option  $\z=0$, i.e.
 \bea
\la{cl}
&&  \G_+ \theta^I= 0 \ , \ \ \ \ \ \ \    \bar \theta^I \G_+ =0 \ , \ \ \ \ \ \ \
(\theta^I)^T\G_- =0 \ , \\
&& \G_\pm \equiv \frac{1}{2}(\pm \Gamma_0+ \G_8)  \ , \ \ \ \ \ \ \
  \G_- \G_+ + \G_+ \G_- =1\ , \ \ \ \ \ \ \  \G^2_\pm=0 \ , \ \ \ \
  (\G_+ \G_-)^2= \G_+ \G_-    \la{tr} \eea
Splitting the bosonic tangent-space  indices into $0,8$ and $p,q=1,2,3,4,5,6,7,9$
we get  from \rf{des}\foot{We have dropped the term with $\omega^{08}_\a$
since this component of the connection vanishes for our direct-product metric.}
\be \la{dere}
D_\alpha\theta^J&=&\partial_\alpha\theta^J
-\frac{1}{2}(\omega_\alpha^{0p}
-{\omega}_\alpha^{8p})\G_-\G_p\theta^J
+\frac{1}{4}\omega^{pq}_\alpha\G_{pq} \theta^J\cr
&+&\frac{1}{2}\epsilon^{JK}e^p_\alpha \G_p \Pi   {\G}_- \theta^K
- \frac{1}{2}\epsilon^{JK}(e^0_\alpha-{ e}^8_\alpha)\Pi \theta^K \ ,
\ee
where
\be  \Pi \equiv \G_{1234} \ ,\  \ \ \ \ \  \G_* = i \G_0 \Pi\ , \ \ \ \ \ \Pi^2 =1  \ .
\la{pii} \ee

The combination entering the quadratic fermionic term \rf{qaq}
becomes
\bea
{\bar\theta}^Ie\llap/{}_\alpha
D_\beta\theta^J&=&
-(e^0_\alpha-{ e}^8_\alpha)
{\bar\theta}^I\G_-\partial_\beta\theta^J
- \frac{1}{4}(e^0_\alpha-{ e}^8_\alpha)\omega_\beta^{pq}
{\bar\theta}^I\G_-\G_{pq}\theta^J\cr
&+&\frac{1}{2}\epsilon^{JK}
(e^0_\alpha-{ e}^8_\alpha)(e^0_\beta-{ e}^8_\beta){\bar\theta}^I\G_-\Pi\theta^K\cr
&-&\frac{1}{2}{\bar\theta}^I{ e^p}_\alpha \G_p \left[
(\omega_\beta^{0q}-{\omega}_\beta^{8q})\G_-\G_q\theta^J
+
\epsilon^{JK}  \G_- e^q_\beta\G_q \Pi  \theta^K
\right] \ . \la{ut}
\eea
Expanding the vielbein and connection near their background values
 in \rf{ljl}  we find for the fermionic kinetic term
\bea
\frac{2\pi}{\sqrt{\lambda}}{\cal L}^{(0)}_{F2}&=&
i(\eta^{\alpha\beta}\delta^{IJ}-\epsilon^{\alpha\beta}{\rm s}^{IJ})\  \big[
{\bar\theta}^I\G_-  { \bar e}^8_\alpha\partial_\beta\theta^J
+\frac{1}{2}\epsilon^{JK}
({\bar  e}^8_\alpha{\bar e}^8_\beta  + {\bar e}^9_\alpha{\bar e}^9_\beta)
  {\bar\theta}^I\G_-  \Pi\theta^K \big] \ ,   \la{iiy} \eea
where (cf. \rf{neq})
\be \la{ppu}
\bar e^8_\a = { \k \ov \sqrt 2}  (  1,- i    ), \ \ \ \ \ \
 \bar e^9_\a ={ \k \ov \sqrt 2} ( 1,i    ) \ , \ \ \ \ \ \ \
 {\bar  e}^8_\alpha{\bar e}^8_\beta  + {\bar e}^9_\alpha{\bar e}^9_\beta
 = - \k^2 \eta_{\a\b} \  ,  \ee
\be \la{relo}
\eta^{\a\b}  {\bar  e}^8_\alpha{\bar e}^8_\beta
= \eta^{\a\b}  {\bar  e}^9_\alpha{\bar e}^9_\beta = - 1   \ , \  \ \ \ \ \ \
\eta^{\a\b}  {\bar  e}^8_\alpha{\bar e}^9_\beta=0 \ ,\ \ \ \ \ \ \
\ep^{\a\b}  {\bar  e}^8_\alpha{\bar e}^9_\beta=  i   \ .
\ee
Thus
\bea
\frac{2\pi}{\sqrt{\lambda}}{\cal L}^{(0)}_{F2}&=&
\frac{\kappa}{\sqrt 2}\bigg[  (1-i)
{\bar\theta}^1\G_-(\partial_1+\partial_0)\theta^1
+ (1+i)
{\bar\theta}^2\G_-(\partial_1-\partial_0)\theta^2\cr
&-& {i \sqrt 2
\kappa}\left({\bar\theta}^1\G_-\Pi\theta^2-{\bar\theta}^2\G_-\Pi\theta^1\right) \bigg]
\equiv  \ha \theta^T K \theta  \  ,  \la{kin}
\ee
where  the kinetic operator in momentum representation is
(we now set $\k=1$)
\be
K=   - {i \sqrt 2 }\left(
\begin{matrix}
(1-i) (q_1+q_0)& - \sqrt  2  \Pi\cr
\sqrt 2 \Pi&   (1+i)   (q_1-q_0)
\end{matrix}
\right) \G_+ \G_-
   \ . \la{liu}
\ee
Here we used that $\bar \theta = \theta^T {\cal C}$,
 ${\cal C}= \G^0=-\G_0$ (see \rf{pr})
and that
$   {\cal C}\G_- = ( \G_- - \G_+)\G_-= - \G_+ \G_-$.

Then  we get for the  propagator  (cf. \rf{few})
\be
K^{-1}=
\frac{ i  }{2 \sqrt 2 (q^2+ 1) }\left(
\begin{matrix}  (1+i)
(q_1-q_0)& \sqrt 2 \Pi\cr
- \sqrt 2 \Pi& (1-i)  (q_1+q_0)
\end{matrix}
\right) \G_+ \G_-
 \ , \ \ \ \ \
K \cdot K^{-1} = \G_+ \G_-   \la{kiu}
\ee
where $\G_+ \G_- = \G_+{\cal C} $,
    $q^2=-q_0^2+q_1^2$.
The propagator  can be written also  in the  following ``covariant''
form:
\be
(K^{-1})^{IJ}=
\frac{ i  }{2 \sqrt 2 (q^2+ 1) }
\big[ (i   \bar e^8_\a  \d^{IJ } +  \bar e^9_\a {\rm  s}^{IJ} ) q^\a
  - \sqrt 2 \Pi \ep^{IJ} \big] \ . \ee
The logarithm of the determinant of $K$ gives the same 1-loop
contribution in  \rf{lji} as found in the $\theta^1= k \theta^2$ gauge.

The  FFB  and FFBB
 interaction vertices are found
from expanding \rf{ut} (multiplied by  ${i} (\eta^{\alpha\beta}\delta^{IJ}
-\epsilon^{\alpha\beta}{\rm s}^{IJ})$
 as in \rf{qaq})  to quadratic order in bosonic fluctuation fields in \rf{jio}
 using the expressions in \rf{ev}--\rf{pip}. Then  the
 Feynman graphs are constructed  using the propagators  \rf{pro}   and \rf{kiu}.
 For example, the  interaction vertices linear in the $S^5$  field $\td x$
 in \rf{jio} are given by
 \bea
&& \frac{2\pi}{\sqrt{\lambda}}{\cal L}_{F2\ \td x}=
 \td x\  {\rm s}^{IJ}   \bar \theta ^I \G_{59} \G_- \theta^J
  + { 1\ov \sqrt 2}  (\del_0   +  i \del_1) \td x\
  {\rm s}^{IJ}  \ep^{JK}   \bar \theta ^I \G_{59} \G_-\Pi  \theta^K   \cr
&& -
 { i\ov 2 \sqrt 2}  (\del_0   +  i   \del_1) \td x \
  \bar \theta ^I \G_{57} \G_-  \theta^I
  -  { 1\ov 2 \sqrt 2}  (\del_0   -  i \del_1) \td x\
    {\rm s}^{IJ}  \bar \theta ^I \G_{57} \G_-   \theta^J
   \ ,  \la{yuu}
 \eea
where we used that a term with $\G_{58}$ similar to the one with $\G_{59}$
  gives   vanishing contribution.




The relevant  4-fermion  terms follow from the general expression in \rf{qqu}.
Using  \rf{pip} ($\bar \omega^{78}_\a =\bar  e^8_\a ,\ \bar \omega^{79}_\a  =-\bar  e^9_\a $)
 first keeping $\bar  e^8_\a, \bar  e^9_\a$ general
and   then using  relations \rf{relo}  we find for the second  term in \rf{qqu}
\begin{eqnarray}
& & \ha (\eta^{\alpha\beta}\delta^{IJ}-\epsilon^{\alpha\beta}{\rm s}^{IJ})
 ({\bar\theta}^K\Gamma^A D_\alpha\theta^K)
({\bar\theta}^I\Gamma_A D_\b \theta^J) = \frac{1}{8 }\big[ -
\bar \theta^K \G^p \G_{-}\G_7 \theta^K \bar \theta^I
\G^p \G_{-}\G_7 \theta^I \cr
&+& i  \ss^{IJ} \epsilon^{JL} \bar \theta^K \G^p \G_{-} \G_7
\theta^K \bar \theta^I \G_p  \G_9 \Pi \G_{-} \theta^L
- i  \ss^{IJ} \epsilon^{KL}  \bar \theta^I \G_p \G_{-} \G_7\theta^J
 \bar \theta^K  \G^p
\G_9 \Pi \G_{-} \theta^L\nonumber\\
&-& \epsilon^{KL} \epsilon^{IM}   \bar \theta^K \G^p
 \G_9 \Pi \G_{-} \theta^L \bar \theta^I \G_p
 \G_9 \Pi \G_{-} \theta^M \big]  \ . \la{yyr}
\end{eqnarray}
The first  term in \rf{qqu} contains two structures:
\begin{eqnarray}
\bar \theta^I e\llap/{}_\alpha {\cal M}^2_{JK}D_\beta
\theta^K=\bar e^8_\alpha \bar \theta^I \G_{-} \ {\cal M}^2_{JK}D_\beta
\theta^K+ \bar e^9_\alpha \bar \theta^I \G_9 {\cal M}^2_{JK}D_\beta
\theta^K\ .
\end{eqnarray}
Computing them  using \rf{oyg},\rf{relo}   we get
\bea
& &(\eta^{\alpha\beta}\delta^{IJ}-\epsilon^{\alpha\beta}{\rm s}^{IJ})
\bar e^8_\alpha \bar \theta^I \G_{-} \ {\cal M}^2_{JK}D_\beta
\theta^K \cr
&=&
\frac{i}{2}  (\eta^{\alpha\beta}\delta^{IJ}-\epsilon^{\alpha\beta}{\rm s}^{IJ})
 \bar e^8_\alpha \epsilon^{LK} \big[\bar \theta^I \G_{-}
 \G^{ij} \theta^J \bar \theta^L \G_{ij} \G_{-}\Pi
\partial_\beta \theta^K-\bar \theta^I \G_{-}
\G^{i'j'} \theta^J \bar \theta^L \G_{i'j'} \G_{-}\Pi
\partial_\beta
\theta^K\big]\nonumber\\
&-&\frac{1}{4} \ss^{IJ}  \epsilon^{LK} \big[\bar \theta^I \G_{-}
 \G^{ij} \theta^J \bar \theta^L \G_{ij} \G_{-} \Pi
\G_{79}\theta^K-\bar \theta^I \G_{-}  \G^{i'j'} \theta^I \bar \theta^L \G_{i'j'} \G_{-} \Pi
\G_{79}\theta^K\big]
\nonumber\\
&+&\frac{i}{4}   \big[\bar \theta^I
\G_{-} \G^{ij} \theta^I \bar \theta^L \G_{ij} \G_{-} \theta^L
 -\bar \theta^I \G_{-}\G^{i'j'}  \theta^I \bar \theta^L \G_{i'j'} \G_{-}
\theta^L\big]\la{om}  \ , \\
 & &  (\eta^{\alpha\beta}\delta^{IJ}-\epsilon^{\alpha\beta}{\rm s}^{IJ})
  \bar e^9_\alpha \bar \theta^I \G_9 {\cal M}^2_{JK}D_\beta \theta^K\cr
  &=&
\frac{1}{2}\ss^{IJ}  \big[\epsilon^{JL}\bar \theta^I \G_9 \G_{-} \
\Pi \G^p \theta^L \bar \theta^K \G_p \G_{-} \G_7 \theta^K \cr &+&
\bar \theta^I \G_9 \epsilon^{LK} \G_{-} \ \G^i \theta^J \bar
\theta^L \G_i \Pi \G_{-} \G_7 \theta^K-  \epsilon^{LK} \bar \theta^I \G_9
 \G_{-} \ \G^{i'} \theta^J \bar \theta^L \G_{i'} \Pi
\G_{-} \G_7 \theta^K\big]\nonumber\\
&-&\frac{i}{2}  \big[ \epsilon^{IL}  \epsilon^{KM}
 \bar \theta^I \G_9  \G_{-} \ \Pi
\G^p \theta^L \bar \theta^K \G_p  \G_9 \Pi \G_{-} \
\theta^M\cr &+& \bar \theta^I \G_9  \G_{-} \G^i \theta^I \bar
\theta^L \G_i  \G_9 \G_{-} \theta^L - \bar \theta^I \G_9  \G_{-}
\G^{i'} \theta^I \bar \theta^L \G_{i'} \G_9 \G_{-} \theta^L\big]\
,  \la{omm}
\end{eqnarray}
where $i,j=1,2,3,4; \  i',j'=5,6,7,9$  and $p=(i,i')$.


\renewcommand{\theequation}{E.\arabic{equation}}
 \setcounter{equation}{0}
\setcounter{section}{1} \setcounter{subsection}{0}

\section*{Appendix E:  Calculation of   2-loop momentum integrals  }

 \subsection*{E.1  Bosonic integrals}



Here we  compute the integral  of $\I_N$ in \rf{inti} that
enters \rf{ttt} and \rf{dii}.
We  split the integrals in the same
way as their integrands  in \rf{inti}
\be\la{in}
 I_{N}= 3 (I_{N, 1}+I_{N,2}+I_{N,3})  \ , \ \ \ \ \ \ \ \ \ \
I_{N,i} = \int
\frac{d^2q_id^2 q_j}{(2\pi)^4}\  \I_{N, i}
 \ . \ee
 Let us
start with $\I_{N,1}$ and
 introduce the  tensor
\begin{eqnarray}
 I_{1}^{\alpha\beta\gamma\delta}&=&\int\frac{d^2q_id^2 q_j d^2
q_k}{(2\pi)^{4}}\ \delta^{(2)} (q_i+q_j+q_k) \
  \frac{ q_i^\alpha q_j^\beta q_i^ \gamma
q_j^\delta \ (q_i^2+q_j^2-q_k^2)^2} {(q_i^2)^2
\left(q_i^2+4\right)
 (q_j^2)^2\left(q_j^2+4\right)    \left(q_k^2+4\right)}\nonumber\\
 &=& { 1 \ov 4} \big[A_1\,\eta^{\alpha\gamma}\eta^{\beta\delta}+
A_2\,(\eta^{\alpha \beta}\eta^{\gamma \delta}+\eta^{\alpha
\delta}\eta^{\beta \gamma})\big]
 \ , \la{fi}
\end{eqnarray}
where we used the  symmetry under
$q_i\leftrightarrow q_j$.\foot{We reinstated  the integral over $q_k$ to make the symmetry
 between $q_i$ and $q_j$ manifest.
 Also, we used the notation $\eta_{\alpha\beta}$
 for the 2d metric.
%
%
The integrand \rf{inti}
 was  already continued to Euclidean space; at the level of the above
analysis  this replaces $\eta_{\a\b}$ with $\delta_{\a\b}$.}
Taking traces over
($\alpha,\gamma)$ and $(\beta,\delta$) we obtain
\begin{eqnarray}
A_1+A_2=\int\frac{d^2q_id^2 q_j d^2
q_k}{(2\pi)^{4}}\ \delta^{(2)} (q_i+q_j+q_k)\  \frac{(q_i^2+q_j^2-q_k^2)^2}
{q_i^2 \left(q_i^2+4\right)
 q_j^2\left(q_j^2+4\right)    \left(q_k^2+4\right)}\ .  \la{hu}
\end{eqnarray}
We  only need that particular  combination of $A_1$ and $A_2$
to compute $I_{N, 1}$.
Expanding the numerator and using various symmetric integration
identities we get from    \rf{hu}
\begin{eqnarray}
I_{N,1}= 4 (A_1+A_2) &=&
4\int\frac{d^2q_id^2q_j}{(2\pi)^{2d}}\Bigg[-
 \frac{4 }{q_i^2 (q_i^2 +4)q_j^2(q_j^2+4)}        \cr &&~~~~
+\frac{2}{q_i^2(q_i^2+4)(q_k^2+4)}+\frac{2}{q_i^2(q_j^2+4)(q_k^2+4)}\cr &&~~~ -\frac{2}{q_i^2 (q_i^2+4)(q_j^2+4)}+
\frac{16}{q_i^2(q_i^2+4)q_j^2 (q_j^2+4)(q_k^2+4)}\Bigg]\cr
&=& 4\int\frac{d^2q_id^2 q_j }{(2\pi)^{4}}\Bigg[-
\frac{4 }{q_i^2 (q_i^2 +4)q_j^2(q_j^2+4)}   \nonumber\\
&+&\frac{1}{q_i^2q_j^2[(q_i+q_j)^2+4]}
+\frac{1}{(q_i^2+4)(q_j^2+4)[(q_i+q_j)^2+4]} \Bigg]\ . \la{onn}
\end{eqnarray}
For  $I_{N,2}$  we proceed in the
same way by starting with  the tensor
\begin{eqnarray}
I_{2}^{\alpha\beta\gamma\delta}&=&\int\frac{d^2q_id^2 q_j d^2
q_k}{(2\pi)^{4}}\ \delta^{(2)}  (q_i+q_j+q_k)\ \frac{q_i^\alpha q_i^\beta q_i
^\gamma q_i^\delta}{q_i^4(q_i^2+4)(q_j^2+2)(q_k^2+2)}\nonumber\\
&=&\frac{1}{8}A_3 \ (\eta^{\alpha\gamma}\eta^{\beta\delta}+\eta^{\alpha
\beta}\eta^{\gamma \delta}+\eta^{\alpha \delta}\eta^{\beta \gamma})\ , \la{jim}
\end{eqnarray}
where $A_3$ is found by taking the trace.  As a result,
\begin{equation}
I_{N,2}=-8 A_3=- 8 \int \frac{d^2q_id^2 q_j}
{(2\pi)^{4}}\frac{1}{(q_i^2+4)(q_j^2+2)[(q_i+q_j)^2+2]} \ . \la{iu}
\end{equation}
For the integral in  the last term  $\I_{N,3}$ in \rf{inti}
 we need to
consider two  tensors associated with the prefactor
\begin{equation}\la{plo}
-(q_{i0}q_{j0}-q_{i1}q_{j1})(q_{i0}q_{k0}-q_{i1}q_{k1})=(q_{i0}q_{j0}-q_{i1}q_{j1})^2+
(q_{i0}q_{j0}-q_{i1}q_{j1})(q_{i0}^2-q_{i1}^2) \ ,
\end{equation}
i.e. one with two  $q_i$'s and  two  $q_j$'s and the other one with
 three  $q_i$'s and  one  $q_j$.
The first one is then similar to $I^{\alpha\beta\gamma\delta}_{1}$ in \rf{fi}
\begin{eqnarray}
I_{3}^{\alpha\beta\gamma\delta}&=& \int \frac{d^2q_id^2 q_j d^2
q_k}{(2\pi)^{4}}   \  \delta^{(2)}  (q_i+q_j+q_k)
\ \frac{  q_i^\alpha q_j^\beta q_i^\gamma q_j^\delta \
[(q_i^2)^2-(q_j^2-q_k^2)^2]}{(q_i^2)^2(q_i^2+4)q_j^2(q_j^2+4)q_k^2(q_k^2+4)}
\nonumber\\
&=&{1 \ov 4} \big[  A_4 \ \eta^{\alpha\gamma}\eta^{\beta\delta}+ A_5 \ (\eta^{\alpha
\beta}\eta^{\gamma \delta}+\eta^{\alpha \delta}\eta^{\beta \gamma}) \big] \ ,  \la{piu}
\end{eqnarray}
where
\bea
A_4+ A_5&=&\int \frac{d^2q_id^2 q_j d^2
q_k}{(2\pi)^{4}}\ \delta^{(2)}  (q_i+q_j+q_k)\
\frac{[(q_i^2)^2-(q_j^2-q_k^2)^2]}{q_i^2(q_i^2+4)(q_j^2+4)q_k^2(q_k^2+4)} \cr
&=&\int\frac{d^2q_id^2 q_j}{(2\pi)^{4}} \Bigg[
\frac{4 }{q_i^2 (q_i^2 +4)q_j^2(q_j^2+4)}
\cr
&&-\frac{1}{q_i^2q_j^2[(q_i + q_j)^2+4]}
+\frac{1}{(q_i^2+4)(q_j^2+4)[(q_i + q_j)^2+4]} \Bigg] \ . \la{opsi}
\end{eqnarray}
The second tensor we need  is
\begin{eqnarray}
\td I_{3}^{\alpha\beta\gamma\delta}&=& \int \frac{d^2q_id^2 q_j d^2
q_k}{(2\pi)^{4}} \  \delta^{(2)}  (q_i+q_j+q_k)\  \frac{q_i^\alpha q_i^\beta
q_i^\gamma q_j^\delta \
[(q_i^2)^2-(q_j^2-q_k^2)^2]}{(q_i^2)^2(q_i^2+4)q_j^2(q_j^2+4)q_k^2(q_k^2+4)}
\nonumber\\
&=&{1 \ov 8}
A_6 \ (\eta^{\alpha\gamma}\eta^{\beta\delta}+ \eta^{\alpha
\beta}\eta^{\gamma \delta}+\eta^{\beta \gamma}\eta^{\alpha \delta})
+
A_7 \ (\eta^{\alpha\gamma}\ep^{\beta\delta}+ \eta^{\alpha
\beta}\ep^{\gamma \delta}+\eta^{\beta \gamma}\ep^{\alpha \delta})\ . \la{yiu}
\end{eqnarray}
The  $A_7$ term is not  contributing
in our case since the combination in  \rf{plo}
is symmetric in $q_i, q_j$ (in fact, $A_7=0$
as one can see by  doing explicitly one of the  two integrals).
Taking traces gives
\begin{eqnarray}
 A_6 &=&\frac{1}{2}\int \frac{d^2q_id^2 q_j d^2 q_k
}{(2\pi)^{4}}\ \delta^{(2)}  (q_i+q_j+q_k)\
\frac{(q_k^2- q_i^2+q_j^2) \
[(q_i^2)^2-(q_j^2-q_k^2)^2]}{q_i^2(q_i^2+4)q_j^2(q_j^2+4)q_k^2(q_k^2+4)}\cr
&=&-\frac{1}{2}\int\frac{d^2q_i d^2q_j }{(2\pi)^{4}}
\Bigg[-
\frac{4 }{q_i^2 (q_i^2 +4)q_j^2(q_j^2+4)}
\nonumber\\
&&\ \ \ +\ \frac{1}{q_i^2q_j^2[(q_i + q_j)^2+4]}
+\frac{1}{(q_i^2+4)(q_j^2+4)[(q_i + q_j)^2+4]} \Bigg]\la{qiu}
\end{eqnarray}
Finally,   we get
\begin{equation}
I_{N,3}=\ 4 \ (A_4+A_5+2 A_6)  \ . \la{gh}
\end{equation}
Summing up the above expressions \rf{onn},\rf{iu} and \rf{gh}
  we obtain for $I_{N}$ in \rf{in}
\begin{eqnarray}
I_{N}&=& 24 \ \int \frac{d^2 q_i d^2 q_j }{(2\pi)^{4}}\Bigg[
\frac{4 }{q_i^2 (q_i^2 +4)q_j^2(q_j^2+4)}
-\frac{1}{q_i^2q_j^2[(q_i+q_j)^2+4]}\nonumber\\
&&\ \ \ -\ \frac{1}{(q_i^2+4)
(q_j^2+2)[(q_i+q_j)^2+2]}
+ \ \frac{1}{(q_i^2+4)(q_j^2+4)[(q_i+q_j)^2+4]}\Bigg]\ .
\label{inta}
\end{eqnarray}
The  integrands on the first line of \rf{inta}
combine into  $ { 8 q_i \cdot q_j \ov q_i^2  q_j^2 (q_i^2 +4)(q_j^2+4)
[(q_i+q_j)^2+4]}$ and the resulting IR finite integral
can be evaluated  using Feynman parametrization.
Alternatively, we may evaluate the two integrals separately
introducing  an IR cutoff $m_0 \to 0$ and using  that
\begin{equation}
\int \frac{d^2 q_i d^2 q_j }{(2\pi)^{4}}
\Big(\frac{1}{q_i^2 + m_0^2 }-\frac{1}{q_i^2+4}\Big)
\Big(\frac{1}{q_j^2+ m^2_0 }-\frac{1}{q_j^2+4}\Big)\ \to \ \
\frac{1}{(4 \pi)^2}\ln^2
({m_0^2\ov 4}) \ ,  \la{joy}
\end{equation}
and also the previously computed expression  \rf{yyy}
for \rf{uul} (see \rf{ip},\rf{uui}), i.e.
\begin{eqnarray}
&& \int \frac{d^2 q_i d^2 q_j
}{(2\pi)^{4}}\frac{1}{(q_i^2+m_0^2)(q_j^2+m_0^2)[(q_i+q_j)^2+4]}=
\frac{1}{(4 \pi)^2}\int_0^1 dx \frac{\ln \frac{4}{m_0^2}+\ln [x
(1-x)]}{4 x (1-x)-m_0^2}\nonumber\\
&&\to  \  \ \frac{1}{4(4
\pi)^2}\bigg[\frac{13}{3}\pi^2+\ln^2(\frac{m_0^2}{4})\bigg] \ .
\end{eqnarray}
The remaining two  integrals in \rf{inta}
are again of the familiar type \rf{ip},\rf{uui}
and are the same as in \rf{ka} and \rf{io}
\begin{equation}
\int \frac{d^2 q_i d^2 q_j
}{(2\pi)^{4}}\frac{1}{(q_i^2+4)(q_j^2+4)[(q_i+q_j)^2+4]}=\frac{1}{4(4
\pi)^2}\int_0^1 dx \frac{\ln [x(1-x)]}{x (1-x)-1}\ ,
 \la{ki}
\end{equation}
\begin{equation}
\int \frac{d^2 q_i d^2 q_j
}{(2\pi)^{4}}\frac{1}{(q_i^2+4)(q_j^2+2)[(q_i+q_j)^2+2]}=\frac{1}{2(4
\pi)^2}\int_0^1 dx \frac{\ln [2x(1-x)]}{2x (1-x)-1} \ .
\la{kap}
\end{equation}
They  are thus expressed in terms of the Catalan constant $\KK$  \rf{cata}
 and a combination of trigamma   values $\td \KK$  \rf{ce}.
Explicitly,
 combining  the values of the above  integrals
  we find  for \rf{inta}
\begin{equation}
I_{N}=-\frac{13}{8}- \frac{24}{\,(4\pi)^2} ( \KK - \td \KK)
\ . \la{hp}
\end{equation}

\subsection*{E.2 Fermionic integrals}

The non-invariant integral in the mixed boson-fermion sector
contains  two different
types of factors.
 The first is (here we  use Euclidean signature and consider the integral
 directly in $d=2$):
\be \la{kapl}
X= (q_{i0}^2 - q_{i1}^2)^2= ( q_i^2 - 2 q_{i1}^2)^2 \ee
 and its expectation value over
$(q_i,q_j)$ symmetric Lorentz-invariant  measure  can be evaluated using that
as in \rf{jim}
$ \langle q^\a_i q^\b_i q^\g_i q^\d _i\rangle = { 1 \ov 8} ( \eta^{\a\b} \eta^{\c\d}  +
\eta^{\a\c} \eta^{\d\b}+
\eta^{\a\d} \eta^{\c\b}) \langle q^4_i\rangle$.
This gives
\be \la{ql}       \langle X_1\rangle = \ha   \langle q^4_i\rangle     \ . \ee
The second combination is
\be  \la{kaa}
Y= (q_{i0}q_{j0} - q_{i1}q_{j1} ) (q_{k0}q_{k0} - q_{k1}q_{k1} )
= (q_{i}^2  - 2 q_{i1}q_{j1} ) (q_{k}^2  - 2q_{k1}q_{k1} )  \ , \ \ \ \ \ \
 q_k = - q_i - q_j \ .
 \ee
Using the $q_i \to q_j$ symmetry of the measure the expectation value of $X_2£$ is the same
as of
\be
Y'
= 2 q_{i}\cdot q_j (q_i^2 + q_{i}\cdot q_j)
  - 4 q_{i1}q_{j1} ( q_i^2 + q_{i}\cdot q_j   )
   - 4 ( q_{i1}^2 + q_{i1} q_{j1})
   + 8  q_{i1} q_{j1} (q_{i1}^2  +  q_{i1} q_{j1}) \ . \ee
Then $\langle Y'\rangle$     can be found by using the same relations as in
\rf{yiu},\rf{piu}
\begin{eqnarray}
 \langle q_i^\alpha q_i^\beta q_i^\gamma q_j^\delta \rangle
 =  { 1 \ov 8} \langle q_i^2 (q_i\cdot q_j) \rangle   (\eta^{\alpha\gamma}\eta^{\beta\delta}+ \eta^{\alpha
\beta}\eta^{\gamma \delta}+\eta^{\beta \gamma}\eta^{\alpha \delta})
\ , \la{yiuj}
\end{eqnarray}
\bea
 \langle q_i^\alpha q_j^\beta q_i^\gamma q_j^\delta \rangle
 = {1 \ov 8} \big[ &&  \langle -2(q_i\cdot q_j)^2  + 3  q_i^2  q_j^2 \rangle
  \ \eta^{\alpha\gamma}\eta^{\beta\delta}\cr
 && +  \langle 2  (q_i\cdot q_j)^2  -  q_i^2  q_j^2 \rangle
  \ (\eta^{\alpha \beta}\eta^{\gamma \delta}+\eta^{\alpha \delta}\eta^{\beta \gamma}) \big]
 \ .  \la{piua}
\end{eqnarray}
As a result,
\be \la{sil}
\langle Y \rangle =  \langle Y'\rangle = \langle  (q_i\cdot q_j)^2  +   q_i^2  q_j^2 \rangle
=  { 1 \ov 4}  \langle (q^2_i+  q^2_j) (q_i + q_j)^2 -  (q^2_i-  q^2_j)^2 \rangle \ .
\ee

Let us now  consider again the
 similar integrals  in $d$ dimensions keeping track of $d$-dependent
 factors.\foot{That may be useful  for finding the coefficient of the
 $1 \ov \ep$ divergences in the fermionic
 sector as in the bosonic sector in \rf{dii},\rf{hjl}.}
  Here we shall use Minkowski signature  and always imply that $q_i+q_j +q_k=0$.
We start with
\be
\int  d^d q_j d^d q_k \, q_k^a q_k^bq_k^cq_k^d \,f(q_j,q_k)=
A\,(\eta^{\a\b}\eta^{\c\d}+\eta^{\a\c}\eta^{\b\d}+\eta^{\a\d}\eta^{\b\c})
\ee
\be
A=\frac{1}{d(d+2)}\int  d^d q_j d^d q_k \, (q_k^2)^2 \,f(q_j,q_k) \ .
\ee
In particular, we find
\be
\int d^d q_j d^d q_k \, (q_{k0}^2+q_{k1}^2)^2
\,f(q_j,q_k)=4A
 \ .
\ee
Let us consider the following combination
\bea
2(q_{i0}q_{j0}+q_{i1}q_{j1}) (q_{k0}^2+q_{k1}^2)&=&
-(q_{i0}^2+q_{i1}^2)(q_{k0}^2+q_{k1}^2)
-(q_{j0}^2+q_{j1}^2)(q_{k0}^2+q_{k1}^2)\\
&&-(q_{i0}q_{k0}+q_{i1}q_{k1}) (q_{k0}^2+q_{k1}^2)
-(q_{j0}q_{k0}+q_{j1}q_{k1}) (q_{k0}^2+q_{k1}^2)
\nonumber
\eea
The reason for this splitting is to maintain the $i\leftrightarrow j$
symmetry. To evaluate its integral we will need
\be
\int d^d q_j d^d q_k \, q_i^a q_i^bq_k^cq_k^d
\,f(q_j,q_k)=A_i\eta^{ab}\eta^{cd}
+B_i(\eta^{ac}\eta^{bd}+\eta^{ad}\eta^{bc})
\ee
Then
\bea
d^2\,A_i+2d\,B_i&=&\int d^d q_j d^d q_k \, q_i^2q_k^2 \,f(q_i,q_k), \cr \ \ \
d\,A_i+d(d+1)\,B_i&=&\int d^d q_j d^d q_k \, (q_i\cdot q_k)^2 \,f(q_i,q_k)\ ,
\eea
and thus
\bea
&&4 B_i=\int d^d q_j d^d q_k \, (q_{i0}^2+q_{i1}^2)
(q_{k0}^2+q_{k1}^2)
\,f(q_j,q_k)\cr
&&=\frac{4}{d(d+1)-2}
\int d^d q_j d^d q_k \,\Big[ (q_i\cdot q_k)^2 -\frac{1}{d}\,
q_i^2\,q_k^2\Big] \,f(q_i,q_k)
\eea
Consider also
\bea
&&\int d^d q_j d^d q_k \, q_j^a q_k^bq_k^cq_k^d
\,f(q_j,q_k)=D_j(\eta^{ab}\eta^{cd}+\eta^{ac}\eta^{bd}+\eta^{ad}\eta^{bc})
\cr
&&\int d^d q_{i,j} d^d q_k \, (q_{i,j0}q_{k0}+q_{i,j1}q_{k1}) (q_{k0}^2+q_{k1}^2)
\,f(q_{i,j},q_k)
\cr
&&=4D_{i,j}=\frac{4}{d(d+2)}\int d^d q_{i,j} d^d q_k \,
(q_{i,j}\cdot q_k)\,q_k^2\,f(q_{i,j},q_k)
\eea
Collecting separate terms we get
\be
2\int d^dq_id^dq_jd^dq_k\delta^d(q_i+q_j+q_k)
(q_{i0}q_{j0}+q_{i1}q_{j1}) (q_{k0}^2+q_{k1}^2)f(q_i,q_j,q_k)\ee
\bea =
\int d^dq_id^dq_jd^dq_k\delta^d(q_i+q_j+q_k)\Big\{
\frac{4}{d(d+1)-2}\Big[(q_i\cdot q_k)^2 +(q_j\cdot q_k)^2 -\frac{1}{d}\,
(q_i^2+q_j^2)\,q_k^2\Big]\cr
+\frac{4}{d(d+2)}(q_i\cdot q_k+q_j\cdot
q_k)q_k^2\Big\}f(q_i,q_j,q_k)
\nonumber
\eea
Using the momentum conservation $q_i+q_j+q_k=0$ we can reorganize
various terms:
\bea
&&(q_i\cdot q_k)^2 +(q_j\cdot q_k)^2+\frac{1}{2}(q_i\cdot q_k+q_j\cdot
q_k)q_k^2 =\frac{1}{2}(q_k^2+2 q_i\cdot q_k)q_i\cdot q_k\cr
&& +\frac{1}{2}(q_k^2+2 q_j\cdot q_k)q_j\cdot q_k
=\frac{1}{2}(q_j^2-q_i^2)q_i\cdot q_k
  +\frac{1}{2}(q_i^2- q_j^2)q_j\cdot q_k=  \frac{1}{2}(q_i^2- q_j^2)^2
\eea
and then
\bea
&&2\int d^dq_id^dq_jd^dq_k\delta^{(d)}(q_i+q_j+q_k)
(q_{i0}q_{j0}+q_{i1}q_{j1}) (q_{k0}^2+q_{k1}^2)f(q_i,q_j,q_k)\cr
&&=\int d^dq_id^dq_jd^dq_k\delta^{(d)}(q_i+q_j+q_k)\cr
&&\Big\{
\frac{2}{d(d+1)-2}\Big[-(q_i^2-q_j^2)^2+\frac{2}{d}\,
(q_i^2+q_j^2)\,q_k^2\Big]-\frac{2(2-d)}{d(d(d+1)-2)}(q_k^2)^2\Big\}f(q_i,q_j,q_k)
\nonumber
\eea
Similarly, we can  compute the integrals
\begin{equation}
I_1=\int d^d q_i d^d q_j d^d q_k \delta^{(d)} (q_i+q_j+q_k)(q_{i0}
q_{j1}+q_{i1} q_{j0})(q_{i0}q_{k1}+q_{i1}q_{k0})f(q_i,q_j)
\end{equation}
\begin{equation}
I_2=\int d^d q_i d^d q_j d^d q_k \delta^{(d)} (q_i+q_j+q_k)(q_{i0}
q_{j1}+q_{i1} q_{j0})(q_{i0}q_{k1}+q_{i1}q_{k0})f(q_i,q_k)
\end{equation}
\begin{equation}
I_3=\int d^d q_i d^d q_j d^d q_k \delta^{(d)} (q_i+q_j+q_k)(q_{i0}
q_{j1}+q_{i1} q_{j0})(q_{i0}q_{k1}+q_{i1}q_{k0})f(q_i,q_j,q_k)
\end{equation}
Let us consider
\begin{eqnarray}
\int d^d q_i d^d q_j d^d q_k \delta^{(d)}(q_i+q_j+q_k) q_i^{a}q_j^b
q_i^c q_j^d f(q_i,q_j)=\frac{1}{4}[A_1 \eta^{a c} \eta^{b d}+ A_2
(\eta^{ab} \eta^{cd}+ \eta^{ad} \eta^{bc})]
\end{eqnarray}
We obtain
\begin{equation}
4 \int d^d q_i d^d q_j d^d q_k \delta^{(d)}(q_i+q_j+q_k) q_i^2 q_j^2
f(q_i,q_j)= A_1 d^2+ 2 A_2 d
\end{equation}
\begin{equation}
\int d^d q_i d^d q_j d^d q_k
\delta^{(d)}(q_i+q_j+q_k)(q_i^2+q_j^2-q_k^2)^2 f(q_i,q_j)=A_1 d+ A_2
d(d+1)
\end{equation}
Then $A_1$ and $A_2$ are
\begin{equation}
A_1=\frac{1}{d^2(d+1)-2d} \int d^d q_i d^d q_j d^d q_k
\delta^{(d)}(q_i+q_j+q_k) [4 (d+1)q_i^2 q_j^2-2
(q_i^2+q_j^2-q_k^2)^2]f(q_i,q_j)
\end{equation}
\begin{equation}
A_2=\frac{1}{d^2(d+1)-2d} \int d^d q_i d^d q_j d^d q_k
\delta^{(d)}(q_i+q_j+q_k) [-4 q_i^2 q_j^2+d
(q_i^2+q_j^2-q_k^2)^2]f(q_i,q_j)
\end{equation}
Also
\begin{eqnarray}
\int d^d q_i d^d q_j d^d q_k \delta^{(d)}(q_i+q_j+q_k) q_i^{\a}q_i^\b
q_i^\c q_j^\d f(q_i,q_j)=\frac{1}{8} A_3 [\eta^{a c} \eta^{\b \d}+
\eta^{\a\b} \eta^{\c\d}+ \eta^{\a\d} \eta^{\b\c}]
\end{eqnarray}
from which one obtains
\begin{equation}
8 \int d^d q_i d^d q_j d^d q_k \delta^{(d)}(q_i+q_j+q_k) q_i^2 q_i q_j
f(q_i,q_j)= d (d+2)A_3
\end{equation}
The integral $I_1$ becomes
\begin{equation}
I_1=-\frac{1}{2}(A_1+A_2+A_3)
\end{equation}
The integral $I_2$ can be written in the same way as $I_1$ with
the formal interchanging $j\leftrightarrow k$ in $A_1$, $A_2$,
$A_3$. The integral $I_3$ is the same as $I_1$.


\end{document}